\documentclass[11pt,preprint,useAMS,usenatbibi,twocolumn]{emulateapj}

\usepackage{graphicx}
\usepackage{epstopdf}
\usepackage{color}
\usepackage{ulem}

\newcommand{\Msol}{\mbox{$M_{\odot}$}}

\shorttitle{ZENS V: Merger Frequencies and Properties}
\shortauthors{A. Pipino et al.} 
\slugcomment{{\sc Accepted for publication in ApJ, 797, 127}}

\begin{document}

\title{The Zurich Environmental Study (ZENS) of Galaxies in
 Groups along \\ the Cosmic Web. V.  Properties and Frequency of Merging \\ Satellites and Centrals in Different Environments \altaffilmark{$\star$}}

\affil{Institute for Astronomy, ETH Zurich, Wolfgang-Paul $I$-Strasse 27, 8093 Zurich, Switzerland}

\email{anna.cibinel@cea.fr}

\author{A. Pipino\altaffilmark{1}, 
A. Cibinel\altaffilmark{1,2}\altaffilmark{$\dagger$},
S. Tacchella\altaffilmark{1},
C. M. Carollo\altaffilmark{1}, 
S. J. Lilly\altaffilmark{1}, 
F. Miniati\altaffilmark{1},\\
J. D. Silverman\altaffilmark{3}, 
J. H. van Gorkom\altaffilmark{4},
A. Finoguenov\altaffilmark{5}
}
 \affil{$^1$Institute for Astronomy, ETH Zurich, Wolfgang-Pauli-Strasse 27, 8093 Zurich, Switzerland}
\affil{$^2$CEA Saclay, DSM/Irfu/S\'{e}rvice d'Astrophysique, Orme des Merisiers, F-91191 Gif-sur-Yvette Cedex, France}
\affil{$^3$Kavli Institute for the Physics and Mathematics of the Universe (WPI), Todai Institutes for Advanced Study, \\The University of Tokyo, 5-1-5 Kashiwanoha, Kashiwa 277-8583, Japan}
\affil{$^4$Department of Astronomy, Columbia University, New York, NY 10027, USA}
\affil{$^5$Department of Physics, University of Helsinki, Gustaf H$\ddot{\mathrm{a}}$llstr$\ddot{\mathrm{o}}$min katu 2a, FI-00014, Helsinki, Finland}
\altaffiltext{$\star$}{Based on observations collected at the European Southern Observatory, La Silla Chile. Program ID 177.A-0680}
\altaffiltext{$\dagger$}{E-mail: \texttt{anna.cibinel@cea.fr}}

\begin{abstract}

We use the {\it Zurich ENvironmental Study (ZENS)} database  to investigate the environmental dependence of the merger fraction
 $\Gamma$ and merging galaxy properties in a sample of $\sim$\,1300 group galaxies with $M>10^{9.2}\Msol$ and $0.05<z<0.0585$. 
 In all galaxy mass bins investigated in our study, we find that  $\Gamma$ decreases by a factor of $\sim 2-3$ in groups with halo masses $M_\mathrm{HALO}>10^{13.5}~M_{\odot}$ relative to less massive systems, indicating a suppression of merger activity in large potential wells.
In the fiducial case of relaxed groups only, we measure a variation $\Delta\Gamma/\Delta \log (M_\mathrm{HALO}) \sim - 0.07$ dex$^{-1}$, which is almost independent of galaxy mass and merger stage. At galaxy masses $>10^{10.2}~M_\odot$, most mergers are dry accretions of quenched satellites onto quenched centrals, leading to a strong increase of $\Gamma$ with decreasing group-centric distance at these mass scales.
Both satellite and central galaxies in these high mass mergers do not differ in color and structural properties from a control sample of nonmerging galaxies of equal mass and rank.  At galaxy masses $<10^{10.2}~M_\odot$ -- where we mostly probe satellite--satellite pairs and mergers between star-forming systems -- close pairs (projected distance $<10-20$ kpc) show instead $\sim2\times$ enhanced (specific) star formation rates and $\sim1.5\times$ larger sizes than similar mass, nonmerging satellites. The increase in both size and SFR leads to similar surface star-formation densities in the merging and control-sample satellite populations. 
\end{abstract}


\section{Introduction}

It is observationally established that galaxy properties, such as star-formation activity or morphology, can be influenced by the surrounding environment \citep[e.g.,][]{oemler74, dressler80, balogh99, baldry06,weinmann06, park07, peng10,Kormendy_Bender_2012, newman12, peng12, wetzel12,carollo14,cibinel13a,cibinel13b}. Among the different processes through which environmental forcing on galaxies can manifest itself, an enhanced rate of galaxy-galaxy interactions or mergers \citep[e.g.,][]{toomre77, feldmann10,kampczyk13} is naturally expected in a $\Lambda$ cold dark matter ($\Lambda$CDM) universe, in which the backbone growth of dark matter halos is well understood in terms of hierarchical accretion of smaller structures within the filamentary cosmic web \citep[e.g.,][]{davis85, pearce99, wechsler02, springel06, maulbetsch07, mcbride09}.  

Together with both internal  (e.g., bar instabilities,  \citealt{Kormendy_1979,combes90,carollo97,carollo99,debattista04,debattista06,Kormendy_Kennicutt_2004}) and environmental secular processes (e.g., ram-pressure stripping or harassment, \citealt{gunn72,Moore1996,Kormendy_Bender_2012}),  mergers can play a substantial role in the evolution of galaxies in high-density environments (\citealt{perez09a,mcintosh08,Kormendy_et_al_2010,kampczyk13}, see also \citealt{Kormendy_2013} for a review on internal versus external processes).
Interactions and mergers between relatively gas-rich galaxies are likely to induce an enhancement in the star-formation rates (SFRs) of participating units or coalesced remnants \citep[e.g.,][]{larson78, mihos94, barnes04, feldmann10, kampczyk13, robotham13} and  are undoubtedly able to alter galaxy morphologies and structural properties \citep[e.g.,][]{toomre72, dekel06, hopkins08, naab99, naab06, feldmann08, feldmann10}.  

A number of studies have investigated galaxy mergers, and their effects on galaxy properties, as a function of galaxy stellar mass, central/satellite rank and environment \citep[e.g.,][]{barton07,perez09a,ellison10, robotham13}. Although there is a general consensus that galaxies interacting with a close companion show a factor of $\sim$2 enhancement in their SFRs \citep[e.g.,][]{larson78, kennicutt87, barton00, lambas03, robaina09, bridge10, robotham13} and in their specific star formation rates (sSFR; \citealt{solalonso06}) relative to their counterparts in isolation, it is still debated whether the strength of the merger-induced star formation varies with environment.
Some of these studies indicate that the enhancement of star formation is predominantly happening in pairs with small ($<20$ kpc) physical separations  \citep{alonso04, depropris05, geller06, perez09a, ellison10, scudder12, patton13}; other works suggest that star formation in merging systems is preferentially enhanced only in the poorest groups \citep{solalonso06} or at relatively low large-scale structure (LSS) densities \citep{perez09b, ellison10, patton11, kampczyk13}. A comprehensive picture is, however, still missing, often because of a lack of disentanglement of the environmental signal from the effect of galaxy mass and because of differences between the adopted environmental definitions in the various studies.

Furthermore, it is also unclear \textit{which} environment is more conducive to galaxy mergers and interactions. The $N$-body simulations of a $\Lambda$CDM universe show that dark matter halo position, orbit and rank in the LSS determine whether and when halos will merge \citep[e.g.,][]{angulo09}. While this indirectly implies a dependence of galaxy mergers and interaction rates on environment, the relative importance of the LSS density, group halo mass, and location within group halos in facilitating galaxy encounters is still unclear. Also debated is whether galaxy mergers can affect both central galaxies and satellite galaxies in the same halos. Semianalytic models of galaxy formation seem to indicate that satellite--satellite mergers are rare and that most mergers involve central galaxies swallowing satellites \citep[e.g.,][]{hatton03, guo11}; currently there is, however, little observational evidence for this.

In this work we utilize the database of the Zurich Environmental Study (ZENS) \citep[][hereafter Paper I]{carollo13b} to make further progress on understanding the dependence of the merger fraction and the properties merging galaxies on (1) galaxy stellar mass, (2) rank of central or satellite within a group halo potential, and (3) local (mass of host group halo and group-centric distance) versus LSS environment.

Specifically we use the environmental measurements of Paper I, the structural measurements for the ZENS galaxies presented in \citet[][hereafter Paper II]{cibinel13a}, and the corresponding photometric measurements presented in \citet[][hereafter Paper III]{cibinel13b} to investigate: (1) how the fraction of galaxy mergers depends, at fixed stellar mass and central/satellite rank, on group halo mass, group-centric distance and large scale structure (over)density; and (2) the structural and star-formation properties of satellites and centrals participating in mergers, relative to nonmerging galaxies of similar stellar mass, rank and environment. 

The paper is organized as follows. In Section \ref{sec:survey}, a brief summary of ZENS is given, with emphasis on the set of environmental, structural and photometric measurements of Papers I, II and III, that we utilize for this study as well as on the definition of the (sub) sample of merging galaxies. The dependence of merger fraction on the mass of the group, group-centric distance and LSS density is presented in Section \ref{sec:env}. We present our results on the properties of merging satellites in Section \ref{sec:satellites} and merging centrals in Section \ref{sec:centrals}. The results are discussed in Section \ref{sec:discussion} and summarized in Section \ref{sec:summ}. Throughout this work we assume $\Omega_m=0.3$, $\Omega_{\Lambda}=0.7$ and $h=0.7$; all magnitudes are in the AB system.

\begin{deluxetable*}{l|ccccc}
\tablewidth{0.9\textwidth}
\tablecaption{Statistics of galaxies in mergers}
\tablehead{ \colhead{All groups} & \colhead{All mergers}  & \colhead{Class 1} & \colhead{Class 2} & \colhead{Class 3} & \colhead{Class 4}  }
Total 	& 162 (74,43)	 & 34 (27,15)	 & 19 (13,6) & 89 (14,2) & 20 (20,20)\\ 
Centrals 	& 32 (15,7)	 & 8 (5,2) &	6 (5,1)	&	14 (1,0)	& 4 (4,4)\\	
Low-mass satellites 	& 81 (38,25)	 & 14 (13,8) & 3 (1,0) & 50 (10,1) & 14 (14,14) \\
\tiny{($[10^{9.2}-10^{10.2}[~M_{\odot}$)} & & & & & \\
High-mass satellites & 49 (21,11) & 12 (9,3) & 10 (7,5) & 25 (3,1) & 2 (2,2)\\
\tiny{($[10^{10.2}-10^{11.7}]~M_{\odot}$)} & & & & & \\

\hline
\multicolumn{6}{c}{Relaxed groups} \\
\hline

Total 	& 123	 & 27	 & 15 & 67 & 14 \\ 	
Centrals 	& 28	 & 8 &	5	&	12	& 3 \\	
Low-mass satellites	& 59	 & 8 & 2 & 40 & 9 \\
\tiny{($[10^{9.2}-10^{10.2}[~M_{\odot}$)} & & & & & \\
High-mass satellites & 36 & 11 & 8 & 15 & 2 \\
\tiny{($[10^{10.2}-10^{11.7}]~M_{\odot}$)} & & & & & 
\tablecomments{\label{t1}Breakdown of the number of central and satellite galaxies per merger type: 
class 1: close pairs with clear signs of mergers, identified as single objects in the 2dFGRS, no confirmation of physical association; 
class 2: same as 1, but physical association confirmed by redshifts found in other surveys (see text); 
class 3: `close kinematic pairs' with projected distances lower than 50 kpc; 
class 4: single galaxies with disturbed morphologies and/or irregular shapes. We refer to these galaxies as `coalesced systems' throughout the paper. 
The numbers in parentheses in the upper half of the Table refer to galaxies in systems with projected separation lower than 20 kpc and 10 kpc (left and right, respectively).
Satellites are further split into two galaxy mass bins ($[10^{9.2}-10^{10.2}[~M_{\odot}$ and $[10^{10.2}-10^{11.7}]~M_{\odot}$), as discussed in Section~\ref{sec:biases}. 
The table shows additionally the statistics for relaxed groups only, where the majority of the mergers occurs.}
\end{deluxetable*}

\section{Dataset} \label{sec:survey}
\subsection{ A brief description of ZENS}

ZENS is based on a sample of 141 galaxy groups extracted from the 2-degrees Field Galaxy Redshift Survey (2dFGRS; \citealt{colless01}), and specifically from the Percolation-Inferred Galaxy Group (2PIGG) catalog \citep{eke04a}. The 141 ZENS groups are an unbiased selection of the \textit{2PIGG} groups in the redshift range 0.05$<z<$0.0585 that have at least five confirmed members brighter than $b_J=19.45$; ZENS group halo masses range from $\sim 10^{12.3} M_{\odot}$ to $\sim 10^{14.8} M_{\odot}$. 

We observed all ZENS groups in the $B$ and $I$ bands with the Wide Field Imager (WFI) camera on the 2.2m telescope at la Silla   (ESO Large Programme 177.A-0680). The details of these observations are provided in Paper II. Briefly, our new imaging  reached a resolution of about 1$^{\prime\prime}$ ($\sim$ 1kpc at the redshift of ZENS) in both bands and, with a total integration time of about 700 sec per group, a depth of $\mu(B )$=27.2mag/arcsec$^{2}$ and $\mu(I )$=25.5 mag/arcsec$^{2}$.
A fully-calibrated set of structural and photometric parameters obtained from these observations is published in the ZENS catalog of Paper I. The derivation of the structural parameters is described in Paper II, and the corresponding photometric measurements are presented in Paper III. Key data products from these works which we utilize in this paper include measurements of galactic sizes ($I$-band half-light radii from single S\'ersic profiles) integrated and spatially resolved $(B-I)$ colors, stellar masses, SFRs and sSFRs. Galaxy stellar masses and (s)SFRs were derived from spectral energy distribution (SED) fitting using the ZEBRA+ code \citep{oesch10}, an unpublished upgraded version of the ZEBRA code of \citet{feldmann06}. Specifically, ZEBRA+ was run using synthetic stellar population models from the \citet{bruzual03} library, with a \citet{chabrier03} initial mass function (IMF).

\subsection{Four Different Measurements of Environment}\label{sec:4environment}

ZENS enables us with a single galaxy sample to investigate how galaxy properties depend, at fixed stellar mass, on four different measurements of environment: $(1)$ the mass of the host group halo, $M_\mathrm{HALO}$, $(2)$ the projected group-centric distance $R$ in units of $R_\mathrm{vir}$, the typical scale radius of the group halo\footnote{$R_\mathrm{vir}$ is defined in Paper I as $R_{200}=\left(\frac{GM_\mathrm{HALO}}{[10H(z)]^2}\right)^{1/3}$, with $H(z)=H_0\sqrt{\Omega_M(1+z)^3+\Omega_{\Lambda}}$ the Hubble constant at the given redshift.}, $(3)$ the LSS (over)density $\delta_\mathrm{LSS}$, and (4) the rank of central or satellite galaxy within the group halo. Paper I presents the computations of these environmental metrics, including a comprehensive set of tests done to assess their robustness; here we only briefly highlight the main steps in their derivation.

The group halo mass $M_\mathrm{HALO}$ was computed from the total group luminosity, integrating the luminosity function to account for the contribution of galaxies below the survey magnitude limit. The luminosity is then transformed into the dark matter halo mass by assuming a mass-to-light ratio calibrated with mock catalogs \citep{eke04b}. Tests done to assess the impact of interlopers and missed galaxies show that the masses of ZENS groups have a statistical uncertainty of about 0.3 dex.

The $\delta_\mathrm{LSS}$ was defined through a fifth-nearest-neighbor algorithm, using however the mass-weighted groups (not the galaxies, as commonly done) as the tracers of the LSS density field.  All galaxy members of a given ZENS group are thus at the same underlying LSS density. This approach reduces the cross-talk which is present, especially at high densities, between LSS density and halo mass/richness (and thus group-centric distance), when using the individual galaxies as Nth-nearest neighbors \citep[see, e.g.,][]{peng10}. 
The typical size of the LSS probed by our estimate of $\delta_\mathrm{LSS}$ is between 1.5 Mpc and 2.5 Mpc (25th and 75th percentiles of the distributions of distances to the 5th nearest group, including ungrouped galaxies, see Paper I).

Both the radial projected position and the central/satellite rank depend on an accurate definition of the central galaxy in each group. We classified the ZENS groups as relaxed and unrelaxed, depending on whether or not a self-consistent solution for a galaxy member to be the central galaxy could be found for that group. To be bona fide centrals, we required galaxies (1) to be consistent with being the most massive member of the group (within the errors associated with our galaxy stellar mass estimates), (2) to be located at a projected distance within 0.5$R_\mathrm{vir}$ from the mass-weighted center of the group, and (3) to have a velocity within one standard deviation from the mean group velocity. In about one-half of the groups, no galaxy member satisfied simultaneously these three conditions. These groups were flagged as unrelaxed and had assigned, as a formal central galaxy, the galaxy member with the highest stellar mass (although either the spatial or the velocity conditions for it to be a genuine central were not satisfied - see Paper I).

\subsection{The Merger Sample}\label{sec:sample}

A total of 162 galaxies with mass above $10^{9.2}~M_{\odot}$ were identified as merging/disturbed systems in the ZENS sample; this includes `interacting galaxies' (\textit{class 1} and \textit{2}), `close kinematic pairs' (\textit{class 3}) and `coalesced systems' (\textit{class 4}) as summarized in Table~\ref{t1}. 
 
Specifically, galaxies in the interacting sample are systems that were identified as single sources in the 2dFGRS (namely they have only one redshift measurement) but for which we found an overlapping companion or clear merger features with another nearby galaxy (with no 2dFGRS spectra) in our WFI imaging. There are 26 such pairs (or triplets, see Figure 41 in Paper II) with a median separation of $\sim$11 kpc. Structural and photometric properties have been measured on each individual galaxy in most of the cases.  In nine of these pairs we found the redshift of both companions in either the Sloan Digital Sky Survey (SDSS) \citep{york00} or the NED\footnote{http://ned.ipac.caltech.edu/} databases (merger flag $=2$ in the ZENS database, \textit{class 2} in this work); this confirmed a physical association of the two galaxies in a merger process. In the remaining cases for which no spectroscopic information is available (merger flags $=1,\, 1.5$ in the ZENS database, \textit{class 1} in this work), the presence of tidal features or disturbed morphologies often supports the merger scenario, although for galaxies flagged as $1$ in the ZENS parent catalog a chance projection may not be excluded. We therefore checked and found that our results do not change if these systems are removed from the analysis. Given the typical separations and the total stellar masses of the 26 pairs we estimate using Equation (10) in \citet{kitzbichler08} that these systems will merge on timescales of a $\sim$300 Myr (median value). In 14 of these 26 pairs, the primary galaxy (i.e., the most massive galaxy participating in the merger) is a central galaxy; in the remaining pairs both participating galaxies are satellites.

Another 89 merging galaxies belong to group `kinematic pairs' (and one triplet: merger flag $=3$ in the parent catalog, \textit{class 3} in this work). These are ZENS group members having a projected distance from another member smaller or equal to the largest separation observed in the `interacting' sample (classes 1 and 2 in Table \ref{t1}, maximal separation of $\sim 50$ kpc) and  a velocity difference that is less than $500~ \rm km \, s^{-1}$. For this subsample, the median projected separation is $\sim 30$ kpc. In a third of these cases a central galaxy is the primary of the merging systems.

The remaining 20 merging systems were identified as such because of a clear multiple galaxy appearance, i.e., a morphology most likely arising from a multiple-galaxy contribution rather than a simple disturbed morphology for a single galaxy (merger flag $=4$) or for having irregular morphologies (morphology type $=5$ in the parent catalog); we interpret these systems (grouped together as \textit{class 4} in Table \ref{t1}) to be at stages where the two progenitor galaxies are no longer fully separable. Only four of these coalesced mergers are central galaxies, the remaining are satellite galaxies. 

In the following and in light of the above discussion, all merger classes will be grouped together when considering galaxies with separation $d<$50 kpc that hence correspond to the global sample of ZENS mergers. We will also discuss the results by restricting the sample to pairs with separations $d<$20 kpc or $d<$10 kpc.

\begin{figure*}
\includegraphics[width=0.9\textwidth]{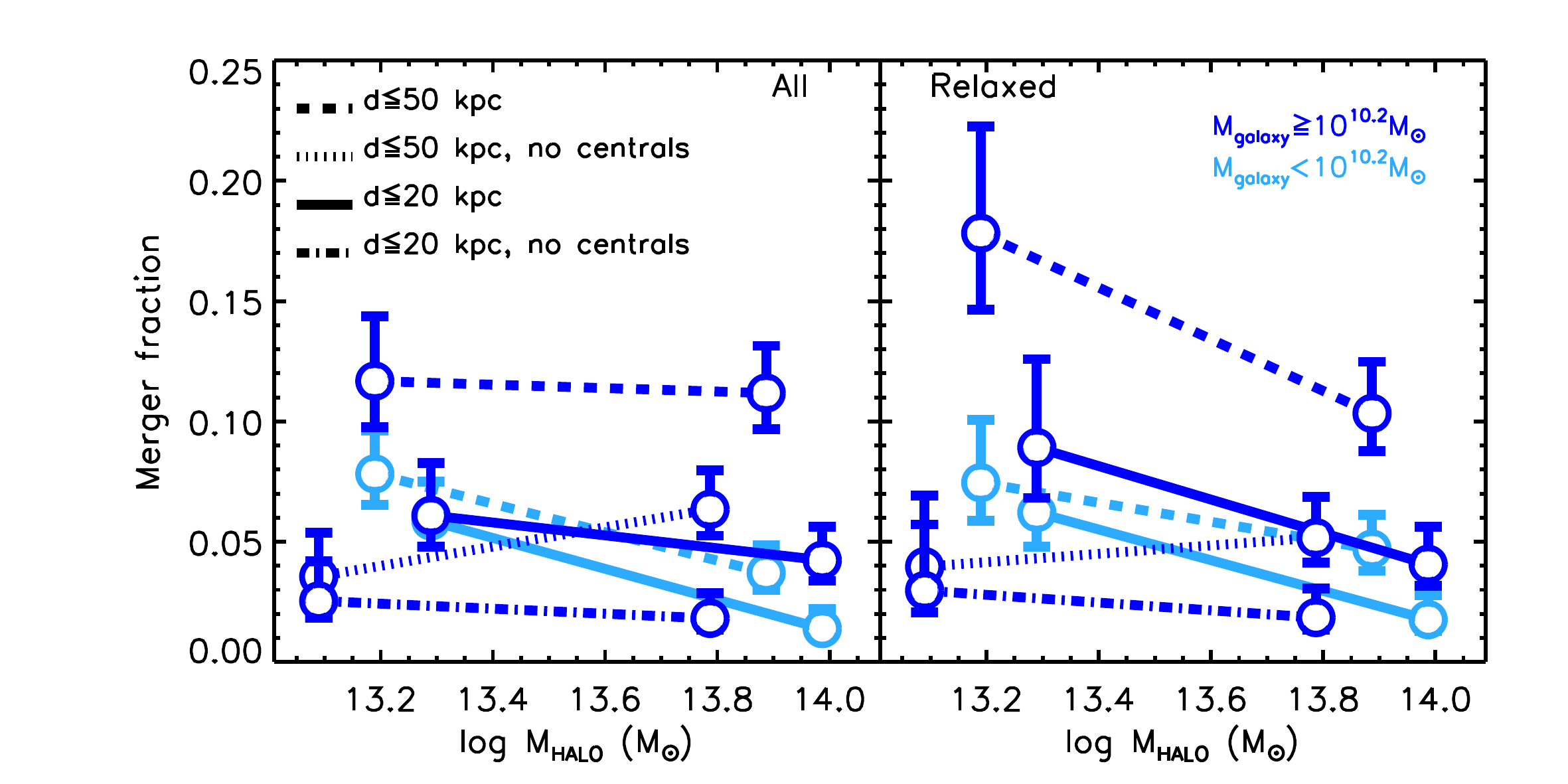}
\caption{Merger fraction as a function of halo mass $M_{\rm HALO}$ in the two bins of galaxy mass here considered (high-mass galaxies are dark blue and low-mass galaxies are light blue). The mass of the primary (most massive) galaxy is used to place each pair in the suitable galaxy mass bin. The left panel shows the results for all ZENS groups, whereas the right panel refers to relaxed groups only. The results for all mergers (i.e., separation $d<50$ kpc) in our sample are shown with dashed lines (dotted when the mergers involving centrals are removed at high galaxy masses), whereas the measured fraction for close ($d<20$ kpc) mergers are shown with solid line (dot-dashed when the mergers involving centrals are removed at high galaxy masses). The position on the x-axis is given by the median environment (e.g. $M_{\rm HALO}$) of $d<50$ kpc merging galaxies (the entire sample) and the points are displaced by $\pm0.1$ dex to improve clarity. The errors on the measured values indicate 1$\sigma$ confidence intervals for a binomial distribution, as calculated in \citet{cameron11} using the beta distribution quantile technique. The merger fraction is enhanced by a factor of $\sim 2-3$ in groups with halo masses $M_\mathrm{HALO}<10^{13.5}~M_{\odot}$ with respect to more massive groups.
\label{Figure_1}}
\end{figure*}

 \subsection{Circumventing Sample Biases: The Galaxy Populations Investigated in This Study}\label{sec:biases}
 
With key galaxy properties depending on galaxy stellar mass \citep[e.g., star formation activity and metallicity,][]{brinchmann04, tremonti04, thomas05, gallazzi05, peng10} and on the central versus satellite rank of a galaxy within its host group (e.g., colors and concentrations; see, e.g., \citealt{weinmann09}, and also our Papers I, II and III), it is important to compare the structural and star formation properties of merging galaxies and nonmerging control-sample galaxies at constant stellar mass {\it and} central versus satellite rank within the host group. Consequently, we will compare $(i)$ merging satellites involved in satellite--satellite mergers, as well as satellite companions in mergers with central galaxies, with nonmerging satellites of similar masses and, $(ii)$ merging centrals with nonmerging centrals of similar mass. 
The corresponding control sample will contain {\it all} nonmerging satellites or centrals with stellar masses within $\pm0.2$dex from the median mass of the merger sample in a given mass bin.

In particular, in the following we split the merger (and control) sample into two mass bins: a `low galaxy mass bin', defined within the range [$10^{9.2}, 10^{10.2}$[ $M_{\odot}$ (median mass: $10^{9.6}~M_{\odot}$), and a `high galaxy mass bin', defined within the range [$10^{10.2},10^{11.7}$] $M_{\odot}$ (median mass: $10^{10.6}~M_{\odot}$ for satellites and $10^{11}~M_\odot$ for centrals). 
 
 The lower mass limit of $10^{9.2}~M_{\odot}$ is the 85$\%$ stellar mass completeness level for our star-forming sample; the corresponding value for quiescent galaxies is $10^{10}~M_{\odot}$ (see Paper II). Thus, our low galaxy mass bin is incomplete for passive galaxies and hence we will  not comment on this population of galaxies. In contrast, a comparative analysis between merging galaxies and nonmerging control-sample galaxies is nevertheless robust because both galaxy populations suffer from identical incompleteness. Furthermore, star-forming galaxies are mass-complete at these mass scales, enabling a sound comparison between merging and nonmerging star-forming galaxies. Note that, because of the sample construction of ZENS (see Papers I and III), in this low-mass bin there are no central merging galaxies and thus our analysis at these mass scales will refer to satellites only.
 
On the other hand, the high galaxy mass bin is also complete for passive galaxies and is substantially populated by both satellite and central galaxies. Therefore, at these higher galaxy masses we are able to explore mergers involving quenched or star-forming galaxies as well as central--satellite or satellite--satellite encounters. In the high-mass bin we have 28 mergers involving a central among the pairs in the sample and four centrals that are in the class of coalesced mergers. The remaining merging systems are, formally, satellite--satellite mergers. Incompleteness in the parent 2dFGRS spectroscopic sample and statistical uncertainties in the group-finding algorithm may of course introduce errors in ranking central and satellite galaxies (see Paper I). As far as we could check, these merging systems involve genuine satellites, at least in the sense that they occur in relaxed groups that, considering the errors on the galaxy stellar mass estimates, clearly host substantially more massive galaxies (identified as the centrals of these groups). Only three mergers occur in unrelaxed groups in which the mass of the primary galaxy is consistent with being the largest in the group (although it has been discarded as central on the basis of velocity constraints; see Paper I).

Finally, the ZENS detection limits (given by the 1$\sigma$ background fluctuations in a uniform area of 1 arcsec$^2$) of $\mu_B$ = 27.2 mag\,arcsec$^{-2}$ and $\mu_I$ = 25.5 mag\,arcsec$^{-2}$ (AB magnitudes) prevent us from detecting (smooth) weak tidal features at lower surface brightnesses \citep[e.g][]{vandokkum05,tal09}. According to simulations \citep{kawata06, feldmann08}, such features are distinctive of either minor mergers or of the late stages of major encounters. They are visible 3--4 Gyr after the mergers if created by the interaction of a bulge-dominated system with a disk-dominated one. In this case, we may only miss mergers that occurred $>3$ Gyr prior to the observation. On the other hand, very low surface brightness features are visible for $\sim$1 Gyr if originated in a merger of two dynamically hot systems \citep{feldmann08}. In this case, strong features quickly fade away, so we may miss somewhat more recent mergers. 

 \subsection{Uncertainties in Color and SFR of Close Pairs}\label{sec:photometricErrors}

In Sections \ref{sec:satellites} and \ref{sec:centrals} we will compare colors and derived properties such as SFR and sSFR of merging and nonmerging galaxies. We hence briefly discuss here the sources of uncertainties on such quantities. We refer the interested  reader to Paper III for a general discussion on errors affecting the parameters derived from the SED fitting and focus instead on the aspects that are more specific to the merger sample.

For mergers in which the two galaxies are sufficiently separated or for the disturbed isolated galaxies (class 4), the main sources of uncertainties are those also affecting the other ZENS galaxies, i.e., background noise and errors in the SED modeling.
Namely, we expect typical errors on the masses, color and sSFR of a given galaxies to be on the order of 0.05 mag, 0.1 dex and 0.2 dex, respectively. We show these average uncertainties for the individual measurements with gray crosses in Figures \ref{Figure_4}-\ref{Figure_7}.

\begin{figure*}
\includegraphics[width=0.9\textwidth]{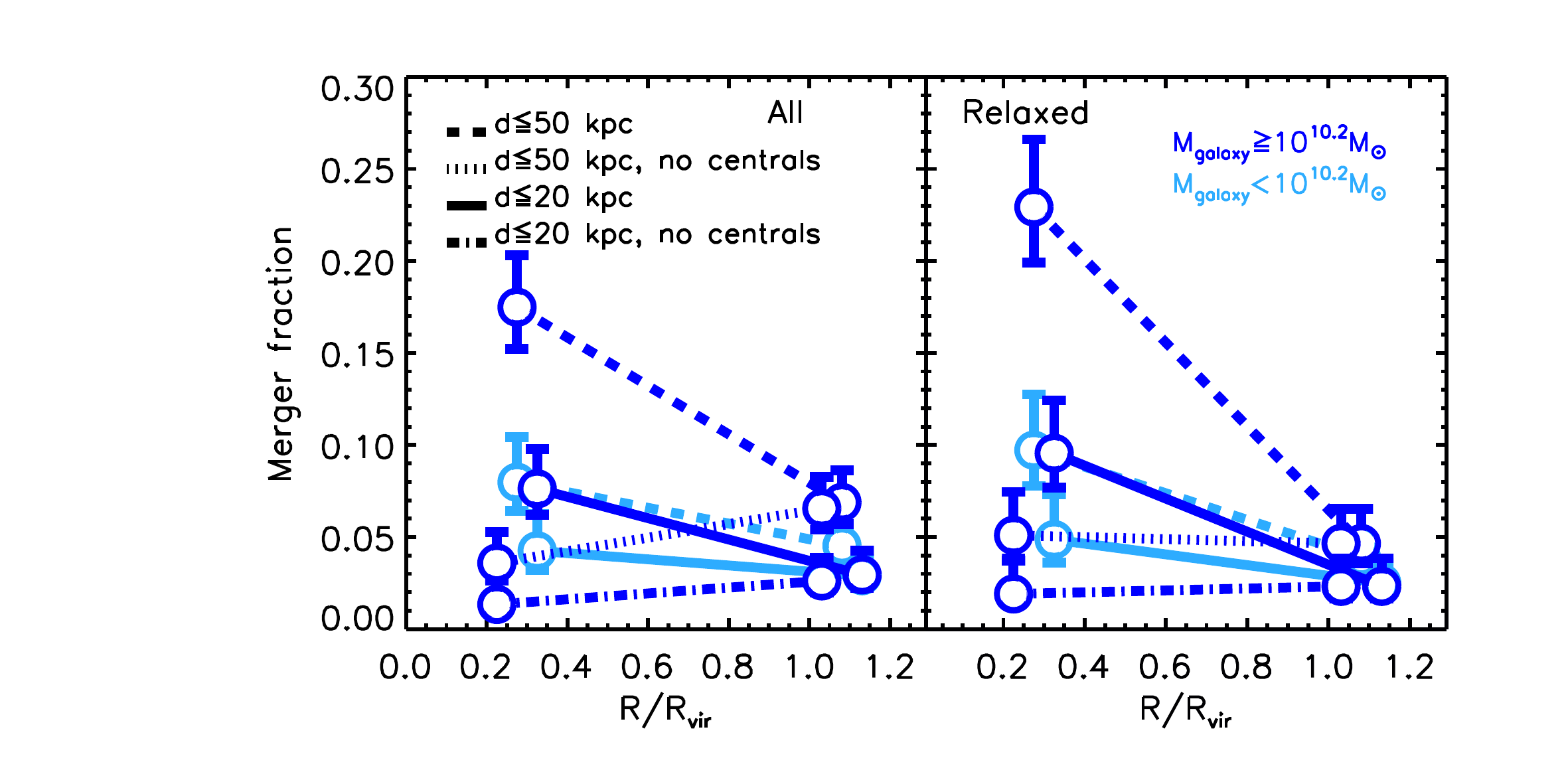}
\caption{Merger fraction as a function of group-centric position $R/R_\mathrm{vir}$ with the same color-coding and line styles as in Figure~\ref{Figure_1}. A decrease in $\Gamma$ of a factor of $\sim$3, moving from the inner ($R < 0.5R_\mathrm{vir}$) group regions toward the group outskirts, is observed at high galaxy masses for all mergers. As shown with the dotted and dash-dotted lines, this increase in the merger fraction with decreasing group-centric distance is however mostly driven by mergers involving (as primary) the central galaxy of the given group, which are at $R/R_\mathrm{vir}=0$ by construction. 
In contrast, our low galaxy mass bin includes only satellite galaxies, implying a mild dependence on group-centric distance of satellite--satellite mergers at galaxy masses in the range $10^{9.2}-10^{10.2}~M_{\odot}$. 
\label{Figure_2}}
\end{figure*}

However, for class 3 mergers or very close pairs almost at coalescence in classes 1 and 2, the blending between the two merging galaxies can introduce further biases in the estimates of magnitudes and colors (and hence masses and SFR).
To test the effects of such galaxy confusion, we generated a set of artificial images consisting of two model galaxies placed at increasingly closer separations, starting from the maximum distance of 50 kpc. 
The simulated galaxies were selected to have magnitudes and sizes within the range observed in our merger sample and the artificial images were processed to reproduce the typical ZENS resolution and noise properties. 
While it is important to probe different combinations of inclinations and steepness of the light profiles for testing the efficiency of recovery of the intrinsic fluxes (see the extensive discussion in Paper II), a dense sampling of all regions of the parameter space is beyond the scope of this test. 
 To bracket the typical observed distributions, we thus created galaxies with either S\'ersic index $n$\,=\,1 or $n$\,=\,4   and  ellipticities between 0 and 0.6.
We then processed these artificial mergers as the real ZENS pairs and compared the fluxes measured within our fiducial  2$\times R_\mathrm{Petrosian}$  aperture (see Paper III for all details of the photometric measurements in ZENS) with those that would be obtained if the galaxy were in isolation\footnote{As discussed in Paper II, biases between the measured and intrinsic (input model) fluxes can arise by a number of observational limitations that are not related to the presence of a nearby companion. For this reason, the input model magnitudes cannot be directly compared with the measured values unless the corrections described in Paper II are applied. A full calibration of the set of models presented here is however beyond the scope of the test. We hence have chosen to compare the fluxes measured on the simulated pairs with those for identical artificial galaxies with no companions. Because these measurements are both affected by the same biases, they enable us to perform a consistent comparison.}.

As a result of this test we found that robust flux estimates can be derived  for pairs with separations down to 8 kpc, resulting in a median magnitude difference from the isolated case smaller than 0.05 mag, i.e. comparable with the photometric uncertainty.
In our sample of class 1 and 2 mergers about 25\,\% (six out of 26) are located at separations smaller than 8 kpc. 
We have  tested our results by excluding these objects with larger uncertainties and found no substantial change with respect to what is discussed in the following for the entire sample.


\section{Merger fractions as a function of halo mass, group-centric distance and LSS density }\label{sec:env}

Let us start by addressing the question of whether and how the galaxy merger fraction $\Gamma$ at a constant galaxy stellar mass depends on any of the environments that we study in ZENS, i.e., group halo mass $M_\mathrm{HALO}$, group-centric distance $R/R_\mathrm{vir}$, and LSS density $\delta_\mathrm{LSS}$. 

Note that, unless specified otherwise, we join together the samples of merging satellites and centrals in the {\it total = central+satellite} merger fractions, independent of galaxy rank within the group potentials. 
Specifically, in each of our galaxy stellar mass and environmental bins, we estimate the merger fraction as $\Gamma=\frac{N_{mergers}}{N_{nonmergers}}|_\mathrm{Mgal}^\mathrm{Env}$, i.e., as the ratio between the number of merging systems in that bin and the number nonmerging galaxies (central plus satellites) in the same galaxy mass and environmental bin. 
For this calculation, we count merging pairs as a single system with the mass assigned by the primary galaxy.


\subsection{Halo Mass}

\begin{figure*}
\includegraphics[width=\textwidth]{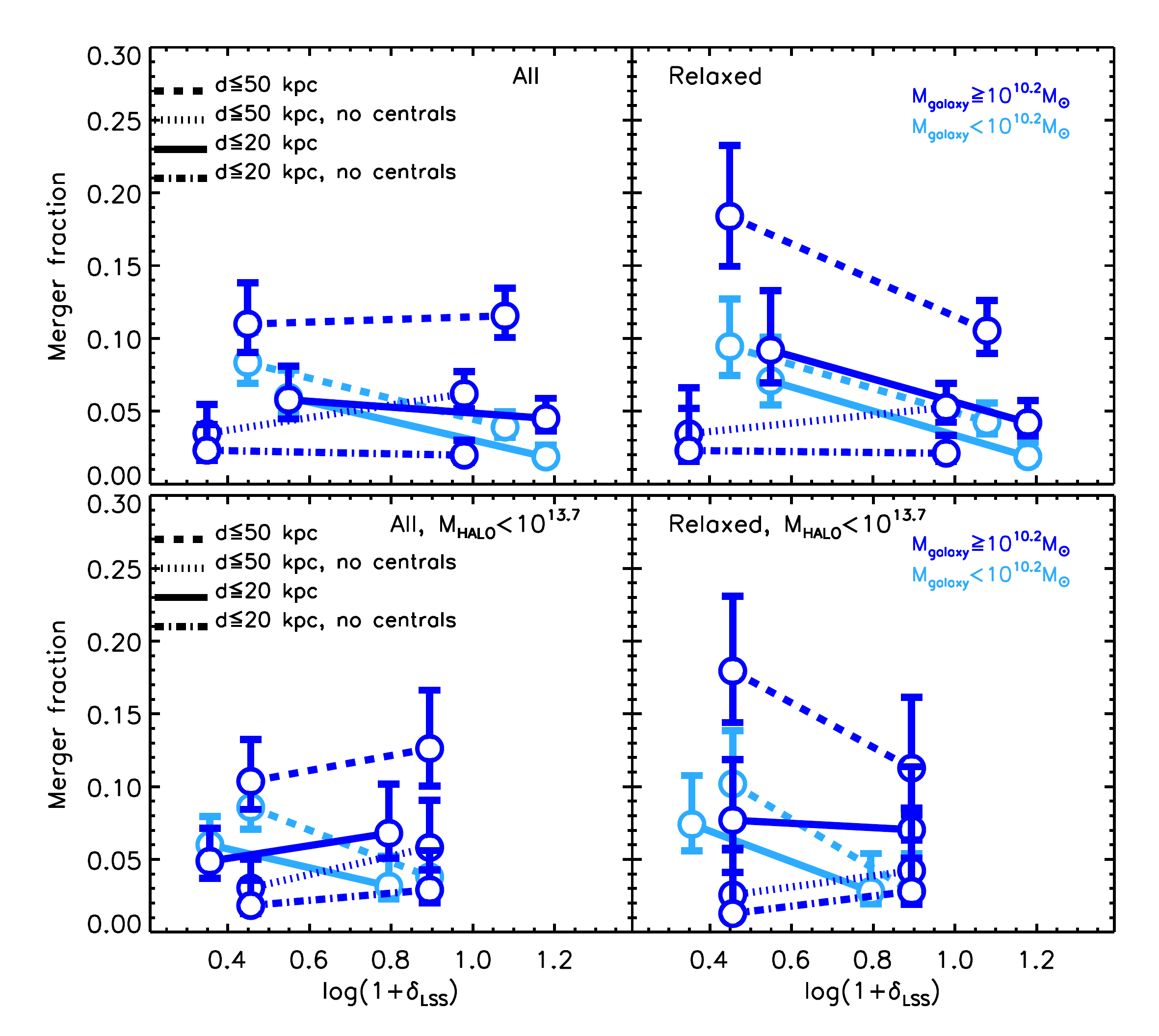}
\caption{Merger fraction as a function of LSS density $\delta_\mathrm{LSS}$ with the same color-coding and line styles as in Figure~\ref{Figure_1}. The upper panels show the results for all group halo masses, whereas the lower panels consider only groups with masses below $10^{13.7}~M_{\odot}$. The left panels refer to groups of any dynamical state and instead in the left panels only relaxed groups are shown. In our low galaxy stellar mass bin, the merger fraction is nominally twice as large in regions of low LSS density. 
At high galaxy stellar masses, no significant trends with LSS density are detected when we limit our study to satellite--satellite mergers, whereas central--satellite mergers are enhanced in low LSS density.
\label{Figure_3}}
\end{figure*}

In Figure~\ref{Figure_1} we show the merger fraction $\Gamma$ as a function of the host halo mass, split into bins of primary galaxy stellar mass (dark blue: high-mass primaries - light blue: low-mass primaries). The results for all mergers (i.e., separation $d<50$ kpc) in our sample are shown with dashed lines (dotted when the mergers involving centrals are removed at high galaxy masses), whereas the measured fraction for close ($d<20$ kpc) mergers are shown with solid lines (dot-dashed when the mergers involving centrals are removed). 

Focusing at first on the entire sample of ZENS groups (left panel in the figure) and $M < 10^{10.2} \Msol$ galaxies, we find that the merger fraction is a factor of $\sim$ 2--3 higher in the low group-mass bin relative to the high group-mass bin. Specifically, $\Gamma$ increases from $\lesssim 3\%$ in groups with $M_\mathrm{HALO}\ge 10^{13.5}~M_{\odot}$, up to $\sim 8\%$ in lower mass groups, with a $\gtrsim3\sigma$ significance at both separations of  $d<20$ kpc and $d<50$ kpc. The decline of the merger fraction between our low and high group-mass bins suggests that low-mass galaxy--galaxy merger activity is virtually fully suppressed in potential wells above mass scales of order $\sim10^{13.5}~M_{\odot}$. No dependence on halo mass is instead observed at galaxy masses $\ge 10^{10.2} \Msol$ when considering all groups; however, if we restrict the sample to relaxed groups only (right panel) a $\ge2\sigma$ decrease in the merger fraction with halo mass is also observed at these high masses. Considering $d<20$ kpc pairs as our reference, we find a variation in the fraction $\Delta\ \Gamma/\Delta \log (M_\mathrm{HALO})=- 0.07\,\mathrm{dex}^{-1}$ for both galaxy mass bins. Similar slopes, within the errors, are also measured at separations of $d<50$ kpc and $d<10$ kpc (not shown), strengthening the inference of the halo mass being the main driver of the change, with no effects due to the stage at which the mergers are observed. 

As discussed in Paper I, the statistical scatter in our group-mass estimates most likely decreases the strength of any measured trend with halo mass; tests performed in Paper I indicate a flattening of a factor of $\sim 1.3$ of the slope of the relationship between merger faction and group halo mass (Paper I). Hence, the increase in the merger fraction from high- to low-mass groups is estimated to be intrinsically $\Delta\Gamma/\Delta \log (M_\mathrm{HALO})\sim-0.1$\,dex$^{-1}$. 

Given the specifics of our sample, removing the mergers that involve a central galaxy (dotted and dot-dashed lines in Figure~\ref{Figure_1}) affects only the high galaxy mass bin.  Because satellite--central mergers make up $\sim 75\%$ of the total number of merger occurrences at these masses,  the net effect is a strong suppression of the merger fraction at any halo mass.
More specifically, at $d<20$ kpc (dot-dashed line) the exclusion of such mergers leaves a flat trend with halo mass. The trend becomes instead positive when considering mergers with all separations (dotted line), owing to the presence of a few massive satellite--satellite mergers in high-mass groups. As discussed above, three such cases occur in unrelaxed groups and the masses of the merging satellites are similar (within the errors) to the masses of the centrals in the same halos. Removing them (by considering only relaxed groups; right panel of the figure) would also produce a flat trend, within errors, of the massive satellite--satellite merger fraction with the host halo mass.
 
Finally, as a consistency check, we note that measured $\Gamma$ values for the entire sample of ZENS mergers are consistent with previously reported merger fractions for low-redshift galaxies of similar masses \citep[e.g.,][]{lotz10, bridge10}. Also, the $\Gamma$ value that we measure in the high galaxy mass bin at high halo masses is in agreement with the results reported by \citet{mcintosh08}, who studied the properties of high-mass ($>10^{10.8}~M_{\odot}$) galaxy pairs with $<30$ kpc separation in SDSS groups above $10^{13.5}~M_{\odot}$; these authors found a merger fraction of $\sim$1--3\% at a halo mass scale of $10^{13.5}~M_{\odot}$, consistent with our result.

\subsection{Group-centric Distance}\label{sec:GroupDistance}

Considering the variation of the merger fraction with distance from the group center in Figure \ref{Figure_2},
the most striking effect we observe is  a strong decrease in $\Gamma$  moving from the inner ($R < 0.5R_\mathrm{vir}$) group regions toward the group outskirts,  if mergers with $M \ge 10^{10.2}\Msol$ are considered. The effect is seen at all separations and at more than the $3\sigma$ level in the full merger sample, where $\Gamma$ varies by a factor of $\sim$3, from $\Gamma\sim 17\%$ to $\Gamma \sim 6\%$.

However, remembering what was already discussed in the previous section regarding the frequency of central--satellite mergers at these high galaxy masses, we stress that the increase in merger fraction toward the group centers is largely driven by mergers involving the central galaxy of the given group, which are at $R/R_\mathrm{vir}=0$ by construction.
In fact, when systems in which the primary is also a central galaxy are excluded (dotted and dot-dashed lines in  Figure~\ref{Figure_2} ) we do not measure any significant change in $\Gamma$ with group-centric distance. This highlights the importance of categorizing galaxies according to their rank within their host-group halos, to avoid incorrect interpretations of the observational signals.

As the only marginal effect of the group dynamical state, we mention that in {\it unrelaxed} groups  the fraction of mergers  with $d>50$ kpc formally increases in the group outskirts when centrals are excluded, with a slope of ${\Delta \Gamma\over \Delta R_\mathrm{vir}} \sim 0.08$ and a $2 \sigma$ significance. One can speculate that the massive satellite--satellite mergers giving rise to this trend have probably just been accreted. Coupled with what was mentioned in the previous section, that their masses are consistent within the errors with that of the nominal central galaxy of their halos, we could identify these mergers as one of the channels that will create the new central galaxy when the group will have relaxed. 

In the low galaxy mass bin, which instead includes only satellite galaxies, we measure  a $\sim 2\sigma$ variation of $\Gamma$ with $R/R_\mathrm{vir}$ when the entire $d<50$ kpc  merger sample is considered  (dashed line) suggesting a  mild dependence on group-centric distance of satellite--satellite mergers at galaxy masses in the range $10^{9.2-10.2}~M_{\odot}$; we note, however, that such an effect is not present if we restrict the sample to only mergers with the closest separations.


\subsection{LSS Density}

Finally, we investigate in Figure~\ref{Figure_3} variations of the merger fraction above and below a threshold value of $\log_{10}(1+\delta_\mathrm{LSS})=0.7$ that separates our two bins of low and high LSS density environment. 
The precise value of 0.7 was chosen to divide the sample of galaxies and mergers in roughly equal numbers while still straddling the transition between high and low LSS densities (see Figure 12 of Paper I).

The upper panels of the Figure present the results with no cut in $M_\mathrm{HALO}$. In the low galaxy stellar mass bin, we observe a decrease of the merger fraction with increasing LSS density,  regardless of the group dynamical state or merger type or separation (taking separations $d=50$ kpc as an example, we measure $\Gamma=4\%$ vs $\Gamma=8\%$ at low and high $\delta_\mathrm{LSS}$, respectively). At high galaxy stellar masses, a similar environmental effect is instead observed only if the sample is limited to relaxed groups and if mergers involving a central galaxy are also considered. On the other hand,  at these masses no significant trends with LSS density are detected when we consider satellite--satellite mergers only. For both the low-mass mergers and the high-mass (including centrals) mergers, we measure ${\Delta \Gamma\over \Delta \log(1+\delta_\mathrm{LSS})} \sim - 0.1$, independent of separation.

These results could however be a consequence of the fact that in our sample $M_\mathrm{HALO}$ and $\delta_\mathrm{LSS}$ start correlating at halo masses above $10^{13.7}~M_{\odot}$ (see Paper I and the trends with halo mass in Figure~\ref{Figure_1}). It is thus important to test whether they still hold when we further restrict our analysis to groups below $10^{13.7}~M_{\odot}$, as shown in the lower panels of Figure~\ref{Figure_3}. In this case, for the low galaxy mass bin, the observed relations remain almost unchanged hinting at a genuine LSS environmental effect, although the significance of the variation of $\Gamma$ in low and high $\delta_\mathrm{LSS}$ decreases to the $<2 \sigma$ level.  At galaxy masses $M \ge 10^{10.2} \Msol$ , the larger errors and the different trends observed when splitting the sample according to the pair separation make the interpretation of the results not straightforward. We nonetheless note that at these high masses a nominal $1\sigma$ decrease of the merger fraction with $\delta_\mathrm{LSS}$ is still also found for relaxed groups with $M_\mathrm{HALO}<10^{13.7}M_{\odot}$ when central--satellites mergers with $d<50$ kpc are considered. 
We highlight this suggested trend as potentially interesting because it would indicate that regions of low LSS density are more conducive to central--satellite mergers.

\begin{figure*}
\begin{center}
\includegraphics[width=0.86\textwidth]{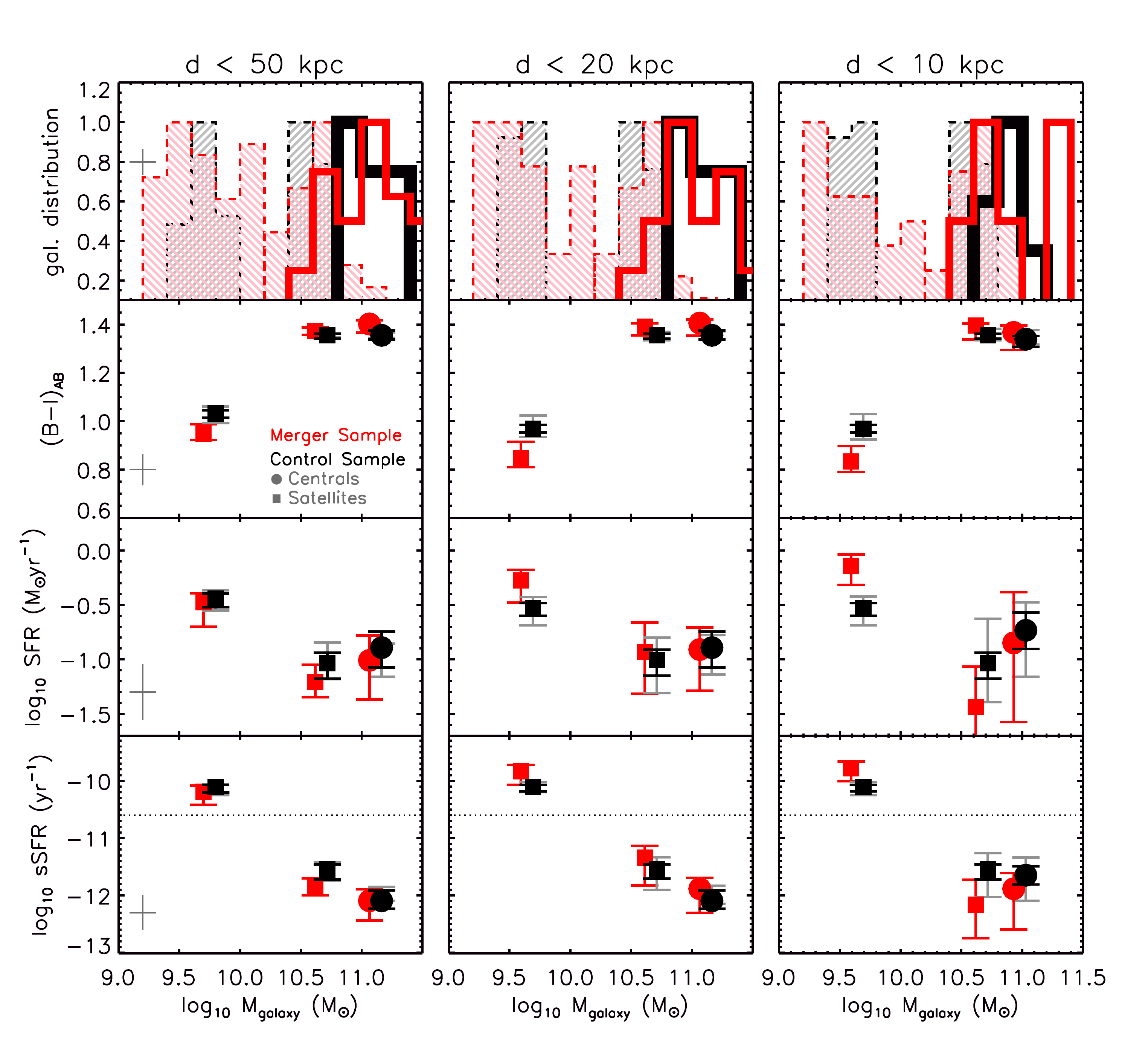}
\caption{Upper row: the normalized mass distributions of galaxies in the merger (red) and control (black) samples. In particular, both merging and control satellites are shown with filled histograms limited by a dashed line, whereas centrals are shown with thicker empty histograms.
The three columns present the results for merging pairs at increasingly smaller separations, as indicated on the top of the panels.
Other rows: as a function of galaxy stellar mass, plotted are the median $(B-I)$ colors, SFR and sSFR of merging satellites (red squares) and centrals (red circles), compared with the corresponding control-sample populations of nonmerging satellites (black squares) and centrals (black circles).  Points for the control samples are shifted by 0.1 dex rightward to increase readability.  The control-sample galaxies are selected to have stellar masses within $\pm 0.2$dex from the median mass of the corresponding merging systems in each broad parent mass bin. Empty symbols indicate bins having $N\leq 5$ galaxies. Error bars (black and red) are derived from the 16th and 84th percentiles of the distributions around the median values. To account for the fact that the size of the control sample can be larger than that of merging galaxies, we show with gray error bars the dispersion around the median  that is obtained in 400 realizations of the control sample in which the number of galaxies is matched to the merger sample. 
In the left panels, the gray crosses on the bottom-left corner indicate the typical uncertainties in the parameters for an individual galaxy.
 The thin dotted line in the bottom panel represents the inverse of the Hubble time, $\tau_\mathrm{Hubble}^{-1} = \log_{10} 1/\rm H_0$, in years; this can be taken to roughly separate passive systems (below the line, with stellar mass doubling timescales longer than $\tau_\mathrm{Hubble}$) and star-forming systems (above the line).
 At galaxy masses of $>10^{10.2}~M_\odot$, satellite and central galaxies are as red and passive as the control sample, and most mergers involve dry accretions of quenched satellites onto quenched centrals. 
At galaxy masses of $<10^{10.2}~M_\odot$, pairs with a projected distance $<$10--20 kpc exhibit $\sim2\times$ enhanced (specific) star-formation rates than noninteracting galaxies.
\label{Figure_4}}
\end{center}
\end{figure*}

\section{Physical Properties of Merging Satellites Relative to Equal-Mass Nonmerging Satellites }\label{sec:satellites}

Let us now turn our attention to the star-formation and structural properties of merging galaxies. For reference, we show in Figure~\ref{Figure_4}, top row, the mass distributions of each subsample of merging galaxies and their nonmerging counterparts that we consider in the analysis. In particular, satellites are shown by shaded histograms, whereas centrals are shown with empty thick line histograms. In this section we focus on the comparison of {\it satellite} galaxies involved in mergers with similar-mass nonmerging satellites (control sample), and we defer the analysis of the properties of merging central galaxies to Section~\ref{sec:centrals}. 

\begin{figure}
\centering
\includegraphics[width=0.45\textwidth,height=0.3\textheight,angle=0]{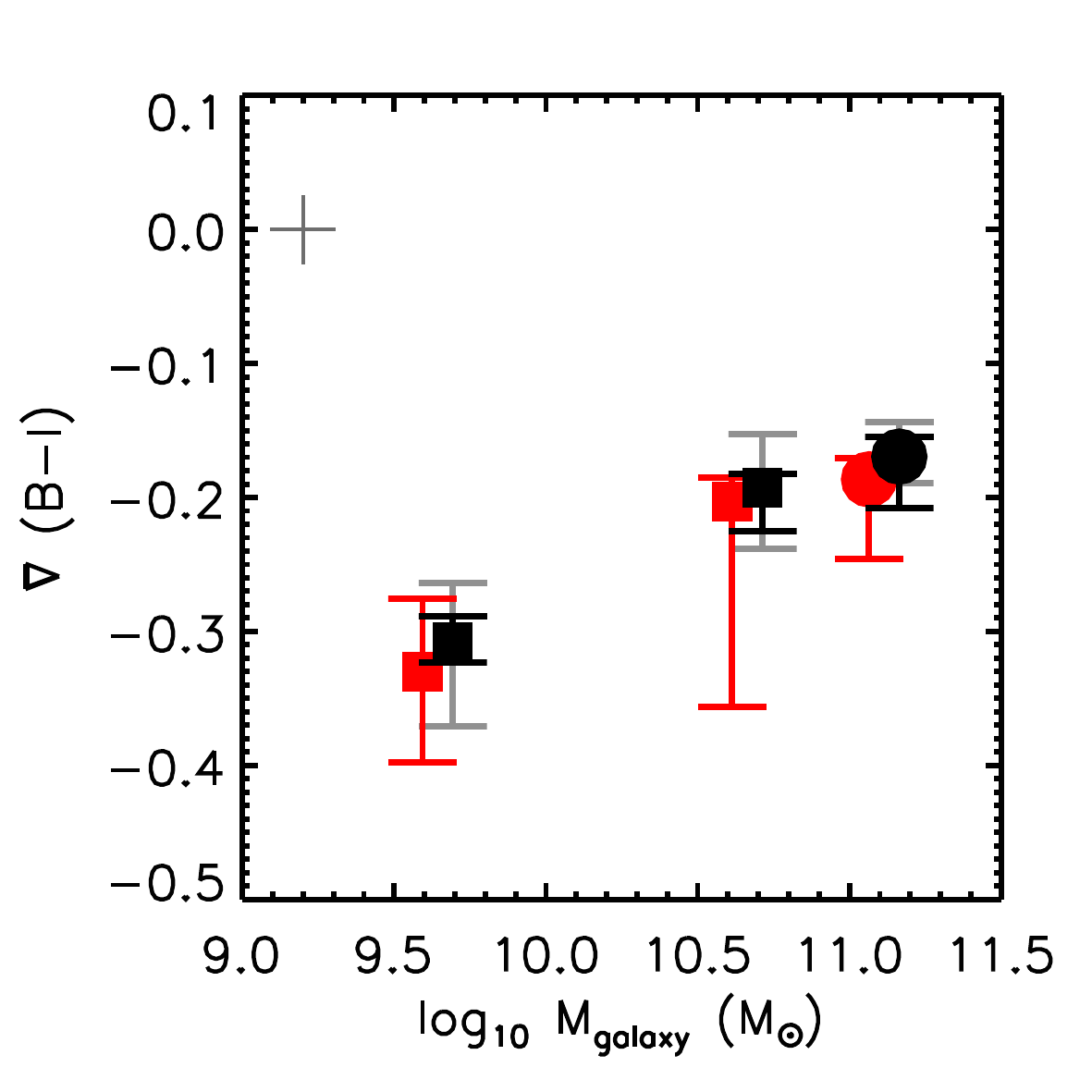}
\caption{Comparison of $(B-I)$ color gradients between merging galaxies and control samples. The results are for $d<20$ kpc mergers and are displayed with the same symbols and colors as in Figure~\ref{Figure_4}.  No noticeable difference in the color profiles of merging and noninteracting galaxies is observed. Note that at high masses all merging galaxies have negative color gradients. At low masses, 10\% of mergers show inverted color gradients, i.e. centrally concentrated star formation, and in 15\% of the low-mass systems the flat color gradients (i.e., $|d\, (B-I)/d\, \log r| < 0.1$ mag) suggest diffuse merger-induced star formation.
\label{Figure_5}}
\end{figure}

We note here that, especially when performing the comparison among satellite galaxies,  the control sample can be substantially more numerous than the merging one. Some of the observed differences between the two populations could thus be a consequence of the lower number statistics in the merging galaxies  and the much higher accuracy of the derived medians in the control sample.
To test how our results would have changed if we had probed the control sample with a smaller number of galaxies,  we calculated the dispersions around the control median values, which are obtained by randomly extracting 400 subsamples that are matched in number to the merger sample.  We show these dispersions in Figures \ref{Figure_4}--\ref{Figure_10} together with the nominal errors derived over the entire sample.

\subsection{Integrated $(B-I)$ Colors} \label{sec:satcolors}

We first investigate the median $(B-I)$ colors of the merging satellites, which are plotted as red squares in the second row of Figure~\ref{Figure_4}. High-mass merging satellites have colors in the range $\sim$1.3--1.5 mag, i.e., as red as those of the nonmerging satellite population of similar galaxy mass, shown with black symbols. The same holds if we restrict the analysis to either mergers with projected separation lower than $20$ kpc (see Figure~\ref{Figure_4}, middle column) or to both a projected separation lower than $10$ kpc (right column) and to coalesced systems (class 4, not shown). 

In the assumption that these red colors are not severely affected by dust reddening and can thus be interpreted with stellar populations models (an assumption that we justify below in Section~\ref{sec:satsfr}, in our analysis of the spectral types and sSFRs of these high-mass systems), such $B-I > 1.3$ mag colors are well modeled with old stellar populations, produced in relatively fast and metal-rich star-formation episodes, followed by stellar passive evolution \citep[e.g.,][]{gunn72, tinsley72, bower92, peletier99, carollo07}. Specifically, using \citet{bruzual03} synthetic stellar population models with exponentially decreasing star-formation histories and a \citet{chabrier03} IMF, typical values for the stellar population of age $A$, star-formation timescale $\tau$ and metallicity $Z$ are, respectively, $A>6$ Gyr, $\tau < 2$ Gyr, and $Z> Z_{\odot}$. An inspection of the spectral types and sSFR of these high galaxy mass merging satellites (discussed in detail in Section \ref{sec:satsfr}) shows that only in one such system the red color [$(B-I)\sim$ 1.4 mag] is due not to passively evolving stellar populations but to dust obscuration of star formation.

Note that, by construction of the ZENS sample, we could have easily seen blue, star-forming mergers; the fact that we do not see them at these galaxy mass scales implies that they are not there in the relatively dense environments probed by ZENS, i.e., galaxy groups with at least five galaxy members and masses above $M_\mathrm{HALO}\sim10^{12.3} M_\odot$.

In the low galaxy mass bin, as discussed in Section \ref{sec:biases}, we are only mass-complete for star-forming galaxies. This is not a limiting factor when comparing properties of merging satellites with properties on nonmerging satellites of similar masses. Overall, low-mass merging satellites have a median color $B-I \sim 1$ mag (left column in Figure~\ref{Figure_4}), slightly bluer but consistent within the errors with that of the control sample. However, if we limit the analysis to those mergers with projected separation lower than $20$ kpc (Figure~\ref{Figure_4}, middle column), the median color of the low-mass merging satellites decreases with respect to that of the full sample and it is $\sim0.2$ mag bluer than mass-matched nonmerging satellites in the control sample. The color difference further increases if we limit the analysis to even smaller separations (right column in Figure~\ref{Figure_4}) or to coalesced galaxies, which have a median color of $(B-I)\sim$ 0.65 mag. It is important to note that these results hold if we remove the irregular galaxies (morphology flag $=5$) from the merger sample and/or if we exclude those close pairs that are likely projection effects (merger flag $=1$). Within the family of Bruzual \& Charlot models above, assuming a metallicity 0.004--0.008, typical of galaxies at these stellar mass scales \citep{gallazzi05}, these low galaxy mass merging satellites are well described by mean ages of $A \approx 5$ Gyr and star-formation timescales of $\tau>7$ Gyr.

A potential source of concern is that the measured effects may be the outcome of slightly different mass distributions between the merger and control sample, even if their median masses are matched. We therefore performed a further test by randomly extracting control samples that matched in number and mass distribution (and, of course, the satellite status) the merger sample in both the low and the high galaxy mass bin. We generated 400 realizations of control-sample galaxies whose mass distribution had a $>90\%$ probability of being drawn from the same distribution of the merger mass distribution according to a Kolmogorov--Smirnov test. For the low-mass bin and in the case of satellites in mergers with $<20$ kpc projected separation, the mean difference in color (namely $<(B-I)_\mathrm{median,mergers}-(B-I)_\mathrm{median,control}>$) between the merger sample and these realizations of the control sample is -0.16 mag, never exceeding -0.06 mag in any single realization, thus confirming our results of a median bluer color in merging galaxies. In the case of low-mass coalesced systems, the mean difference in color over 400 realizations is --0.26 mag (and limited between --0.15 and --0.35 mag). Similarly reassuring conclusions apply to the satellites in the high-mass bin, as well as to the centrals (see Section~\ref{sec:centrals}).

\subsection{SFRs and sSFRs}\label{sec:satsfr}

The availability of (s)SFRs and stellar masses, from fits to the UV-to-NIR SEDs of ZENS galaxies (from Paper III), enables us to get a more precise picture of the origin of the bluer-than-normal colors at low stellar masses and normally red colors of merging satellites relative to nonmerging  satellites at high masses. 

\begin{figure*}
\begin{center}
\includegraphics[width=0.8\textwidth]{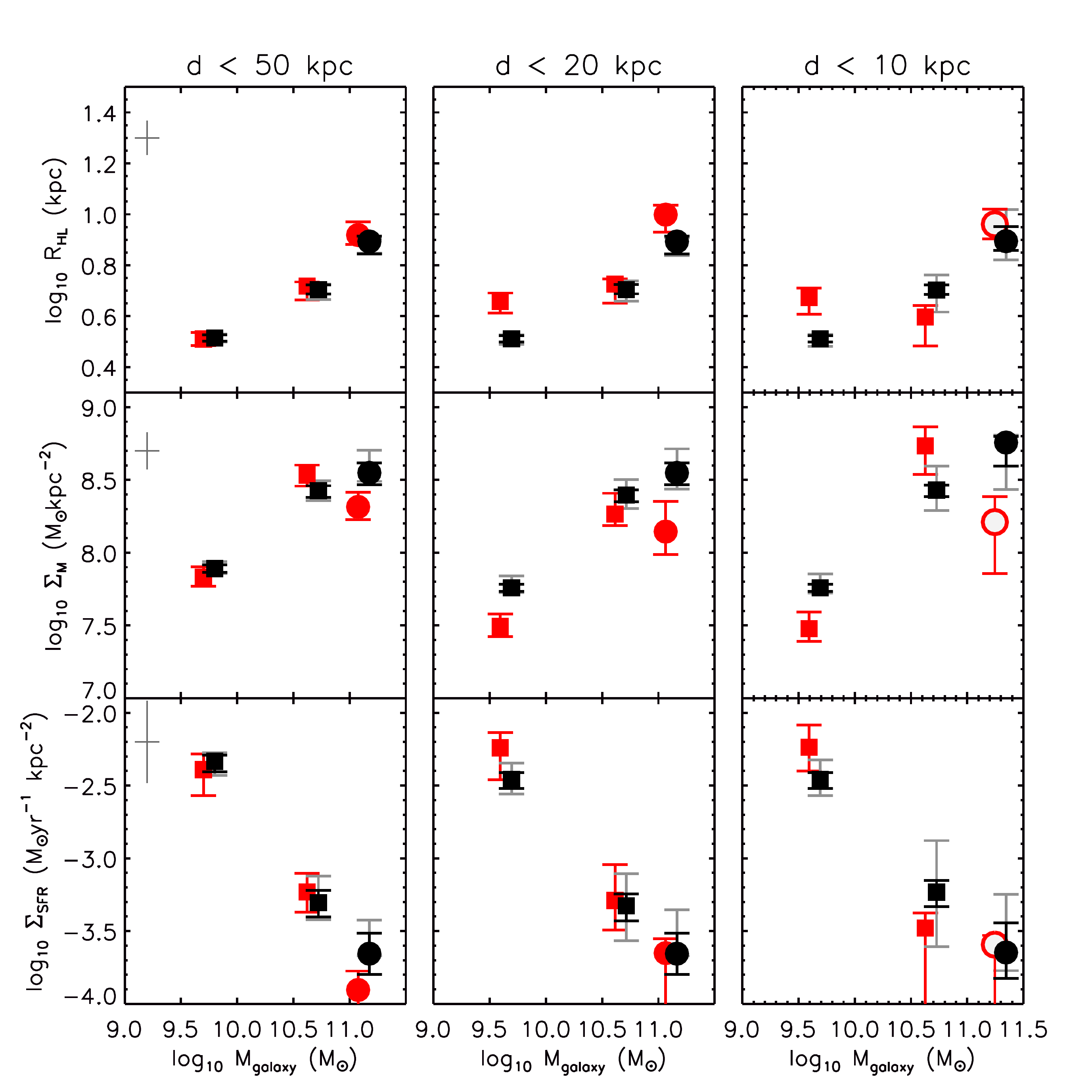}
\caption{
As a function of galaxy stellar mass, we plot the median $I$-band half-light radius $R_\mathrm{HL}$ (top panel), surface mass density $\Sigma_M$ (middle panel), and surface star-formation density $\Sigma_{SFR}$ (bottom panel) of merging satellites and centrals, compared with the corresponding control-sample populations of nonmerging satellites and centrals. Symbols and colors are as in Figure \ref{Figure_4}.  In the high galaxy mass bin, merging satellites have sizes and surface mass densities that are comparable to those of nonmerging satellites of similar mass. Lower mass galaxies have instead $\sim1.5\times$ larger sizes than similar-mass, nonmerging satellites, resulting in lower surface mass density but comparable surface star-formation densities as a consequence of the SFR enhancement (see Figure \ref{Figure_4}).
\label{Figure_6}}
\end{center}
\end{figure*}

The third and bottom rows of Figure~\ref{Figure_4} show the SFR and sSFR versus galaxy stellar mass relations, respectively. In the high galaxy mass bin, consistent with their $(B-I)$ color discussed above, merging satellites have median SFR and sSFR values typical of quiescent populations ($\log \, \rm sSFR/yr \ll-11$), confirming their nature as red-and-dead systems participating in gas-less `dry' mergers. If anything, they are even marginally `more quiescent' than their control-sample nonmerging relatives. We do not find substantial differences if we consider either the closest (separation $<20$ kpc, middle and right columns of Figure~\ref{Figure_4}) or the coalesced systems (not shown).

Close ($<20$ kpc in projection) merging satellites in the low galaxy mass bin, in contrast, show SFR values slightly below $\sim$ 1 $M_\odot$ yr$^{-1}$, and median sSFR $\sim$ 0.7/Gyr which are a factor of $\sim$ 2--3 enhanced above the median rates of the control sample of nonmerging satellites at similar galaxy masses. The difference becomes even more significant if one limits the sample to pairs closer than $10$ kpc in projection (right panel). Similar values are obtained if we limit the analysis to coalesced mergers only.

\begin{figure*}
\begin{center}
\includegraphics[width=0.8\textwidth]{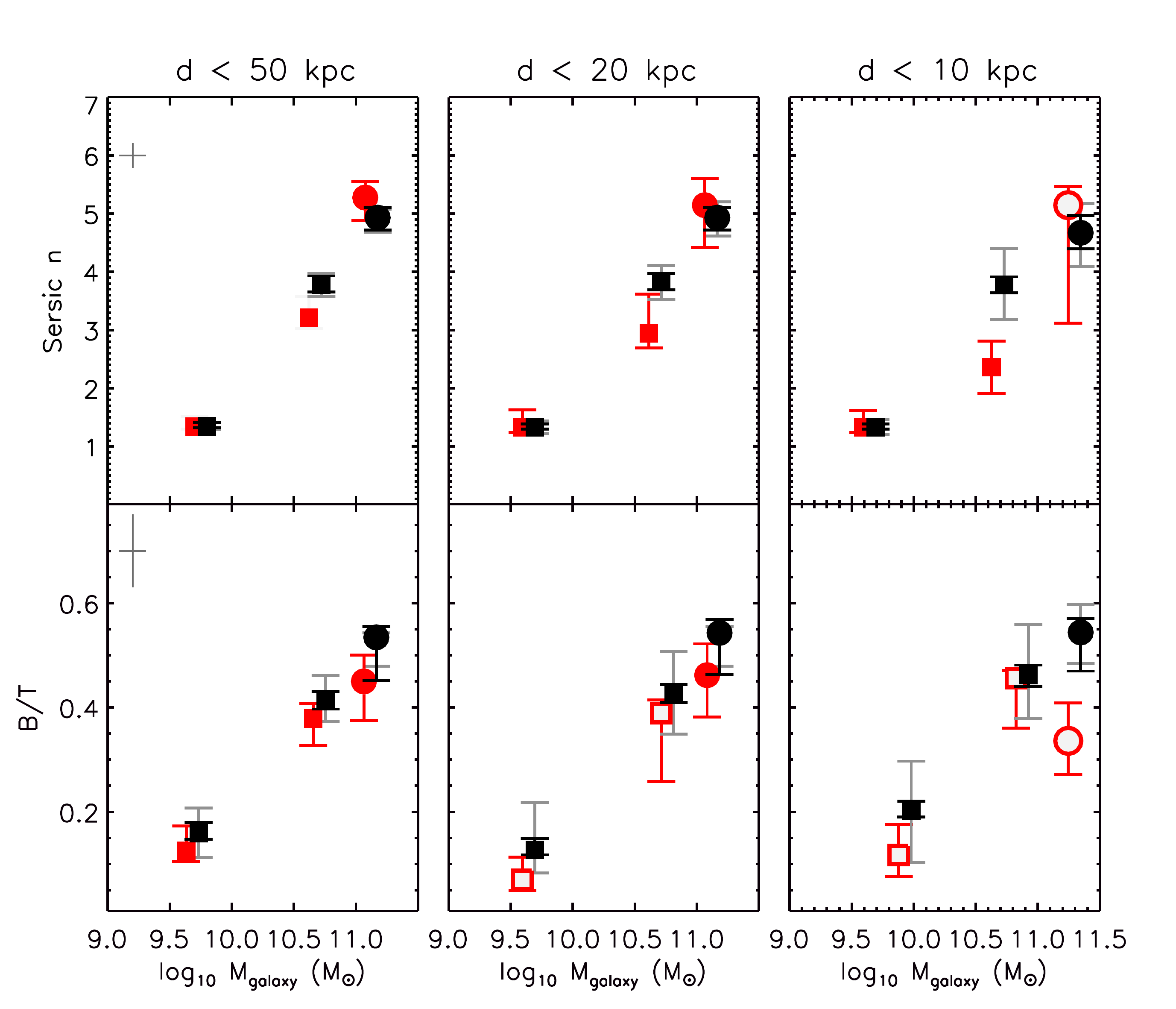}
\caption{Plotted are the median $I$-band S\'{e}rsic index $n$ (top panel) and $I$-band bulge-to-total ratio $B/T$ (bottom panel) of merging satellites  and centrals as a function of galaxy stellar mass, compared with the corresponding control-sample populations of nonmerging satellites and centrals. Symbols and colors are as in Figure ~\ref{Figure_4}.  We find low $I$-band S\'{e}rsic indices in low-mass galaxies involved in mergers and no significant differences for the $I$-band bulge-to-total ratios for mergers and non-mergers at all galaxy masses.
\label{Figure_7}}
\end{center}
\end{figure*}

As before, we performed a further test by randomly extracting control samples that matched in number and mass distribution (and, of course, the central versus satellite status) the merger sample in both the low- and the high-mass bin. In the low-mass bin and in the case of satellites in mergers with $<20$ kpc projected separation, the mean difference in $\log$SFR over these realizations is 0.38 dex, never becoming lower than +0.12 dex, thus confirming our results. 

The bluer colors of low-mass merging satellites relative to nonmerging satellites of similar mass presented in Section \ref{sec:satcolors} are thus indicative of a merger-triggered enhancement of their SFRs, occurring only when the separation between the two galaxies is relatively small. The corresponding mass doubling timescales are about $\sim 1.5$ Gyr, i.e., a factor $\sim$ 3--5 longer than merging timescales inferred from simulations \citep[e.g.,][]{cox06, kitzbichler08}.

\subsection{$(B-I)$ Color Gradients} \label{sec:gradcolors}

Having established that close mergers induce a blueing of the low-mass galaxies, it is interesting to see if the induced star formation is a burst in the center of the galaxies or if it is distributed throughout the galaxy and if it results in distinct color profiles with respect to the noninteracting galaxy sample. For a quantitative analysis, we make use of the radial color gradients derived from the analytical fits of the galaxy light profiles (Paper II) and we compare in Figure~\ref{Figure_5} the gradients for $d<20$ kpc mergers with those in the control sample. 

We do not find any significant difference between the color profiles of satellite galaxies involved in close mergers and those in the control sample. In absolute terms, in high-mass merging galaxies only negative color gradients are found. At low masses, where merging satellites are bluer than the control sample, only a small fraction (10\%) of our merger sample shows inverted color gradients, i.e., blue cores, indicative of centrally concentrated star formation, relative to redder galaxy outskirts. Another 15\% of the low-mass merging systems has instead rather flat (i.e., $|d\, (B-I)/d\, \log r| < 0.1$ mag) color gradients, suggesting that the merger-induced star formation affects the whole system \citep[see also][]{knapen09}. We do not find, however, a correlation between the amount of induced star formation (or the size) and the presence of an inverted or flatter gradient.

\subsection{Sizes, Stellar Mass and SFR surface densities}\label{sec:satsizes}

In order to infer the presence of any merger-induced structural perturbations and check whether merging satellites may sustain higher local SFRs, we also compare in Figure~\ref{Figure_6} the $I$-band half-light radii from analytical single S\'ersic fits (top panel), the stellar mass surface densities (middle panel) and the surface star-formation densities (bottom panel) of merging satellites with the control sample of nonmerging satellites of matched stellar masses \footnote{It was not possible to obtain reliable measurements of the radii for 12 merging galaxies. They are therefore not included in any panel of Figure~\ref{Figure_6}.}.

In the high galaxy mass bin, merging satellites with $d<50$ kpc have sizes ($\log_{10} \, (R_{HL}/$kpc$) \sim 0.7$) and surface mass densities ($\log_{10} \, \Sigma_M \sim 8.5~M_{\odot}$kpc$^{-2}$) that are comparable to those of nonmerging satellites of similar mass. An inversion of the \citet{schmidt59} law using \citet{kennicutt98} and the measured median $\Sigma_M $ value result in a median gas-to-stars ratio in these galaxies $\mu=m_\mathrm{gas}/m_\mathrm{stars} \ll0.1$ \citep[see also][]{ellison10}.

\begin{figure*}
\centering
\includegraphics[width=0.9\textwidth]{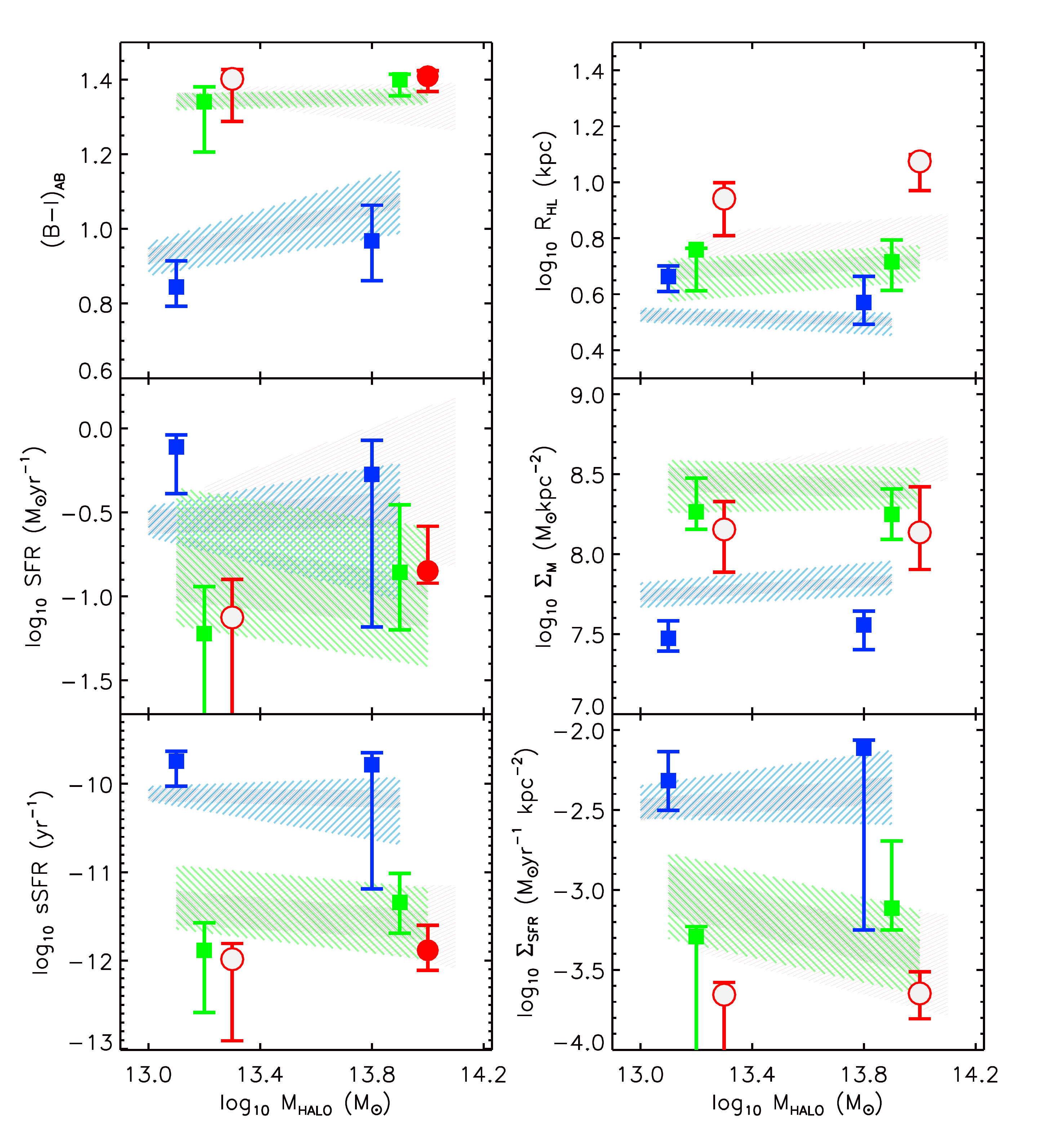}
\caption{Comparison of properties for merging galaxies with the control samples at fixed $M_\mathrm{HALO}$, in two bins above and below $M_\mathrm{HALO}=10^{13.5}~M_{\odot}$. The figure shows only
mergers with projected separation $d$\,$<$\,20 kpc. Low-mass satellites are shown in blue, higher mass satellites are in green, and centrals are in red. Symbols with error bars correspond to the medians obtained on the merging sample, and the shaded areas are the medians and dispersions for the control sample.  Darker shades show the nominal dispersions obtained on the entire control sample. To account for the fact that the size of the control sample can be larger than that of the merging galaxies, we indicate with lighter areas the dispersions around the medians  that are obtained in 400 realizations of the control sample in which the number of galaxies is matched to the merger sample. The differences in the two dispersion estimates for the central galaxies are small, therefore light-shaded areas are shown for satellites only to improve readability. Empty symbols indicate bins having $N\leq 5$ galaxies.
\label{Figure_8}}
\end{figure*}

In the low galaxy mass bin, structural differences between (star-forming) merger galaxies and the control sample are only visible for the closest or coalescing systems (middle and right panels of Figure~\ref{Figure_6}, respectively), which are typically $\sim 1.3$ times larger than their nonmerging counterparts. With a typical (median) $\log_{10} \, \Sigma_M \sim 7.4~M_{\odot} $kpc$^{-2}$, these low-mass merging satellites have a factor of $\sim 2$ lower surface mass densities than nonmerging satellites of similar stellar mass. For them, the inferred median gas-to-stars ratio is $\mu\approx0.5$, to be compared with the $\mu\approx0.3$ value inferred for the nonmerging satellite population at these galaxy masses. It is remarkable, however, that in such low-mass merging satellites, their larger sizes and enhanced sSFRs compensate for each other to result in similar median surface star-formation densities ($\Sigma_\mathrm{SFR} \sim 10^{-2.3}~M_{\odot}$ yr$^{-1}$ kpc$^{-2}$) in both merging and control samples.

\subsection{Morphologies}\label{sec:satmorph}
 
It is now interesting to see whether the changes in the star-formation properties we discussed in the previous sections are followed by significant morphological transformations. We quantify galaxy morphology in terms of the global S\'ersic index $n$ and bulge-to-total ratio B/T in Figure~\ref{Figure_7}.

We find low $I$-band S\'{e}rsic indices in low mass galaxies involved in mergers irrespective of the separation. More quantitatively, and considering the sample of mergers with $d<20$ kpc separation, the median S\'{e}rsic index in merging satellites is $1.33$ (1.26--1.63 are the 16th and 84th percentiles, respectively). The median value for the control sample is 1.32, hence showing no difference to be associated to the mergers. At higher galaxy masses, the S\'{e}rsic indices increase in both the mergers and the control sample, again showing no significant differences for separations $d\ge20$ kpc. The closest ($d\le 10$ kpc) merging satellites have instead significantly lower S\'{e}rsic indices than the control sample. However, there is no S\'{e}rsic index determination available in the parent ZENS catalog for five of 11\footnote{In the low-mass bin, the S\'{e}rsic index is available for 23 of 25 of the merging galaxies with $d<10$ kpc.} of the merging galaxies. Therefore, the suggested trend should be confirmed with larger samples. 

Likewise, no differences are found for the $I$-band bulge-to-total ratios\footnote{Available in the parent catalog only for systems where the bulge/disk decomposition could be performed and that are not best fit by an elliptical morphology.} (see Figure~\ref{Figure_7}, bottom panel) at all galaxy masses and separations:
we measure a median value of $\sim$0.1 in low-mass merging satellites and of $\sim0.4$ in high-mass ones, in agreement with the median values for the control sample. 
This consistency is also reflected in the morphological mix of noninteracting and merging galaxies.
Under the reasonable assumption that incompleteness in the sample of (mostly quiescent) early-type\footnote{Morphological classes 1 and 2; see Paper II.} galaxies affects equally the merging and the control sample, for the low-mass bin we estimate that the fraction of early-type satellites involved in all mergers is 10/81, which is very close to the 13\% featured by the control sample. Similarly, at high galaxy masses where the sample is complete, the fraction of early-type bulge-dominated systems is 11/49, which is 22\% as in the control sample. 

In conclusion, we do not find any significant structural difference that might be induced by the merger process even at the low galaxy mass scale and smallest separations, where we instead observe an enhancement of the SFR and larger radii in the merging population.

\subsection{Environmental Effects}\label{sec:EnvEffects}

In the spirit of ZENS, we now investigate if any of the merger-induced changes in the stellar properties and their trends with galaxy mass can be ascribed to or enhanced by a particular environment. For this analysis, we limit the discussion to mergers with projected separation $< 20$ kpc, where the merger-induced variations in colors, radii and SFRs become significant.

\begin{figure*}
\centering
\includegraphics[width=0.9\textwidth]{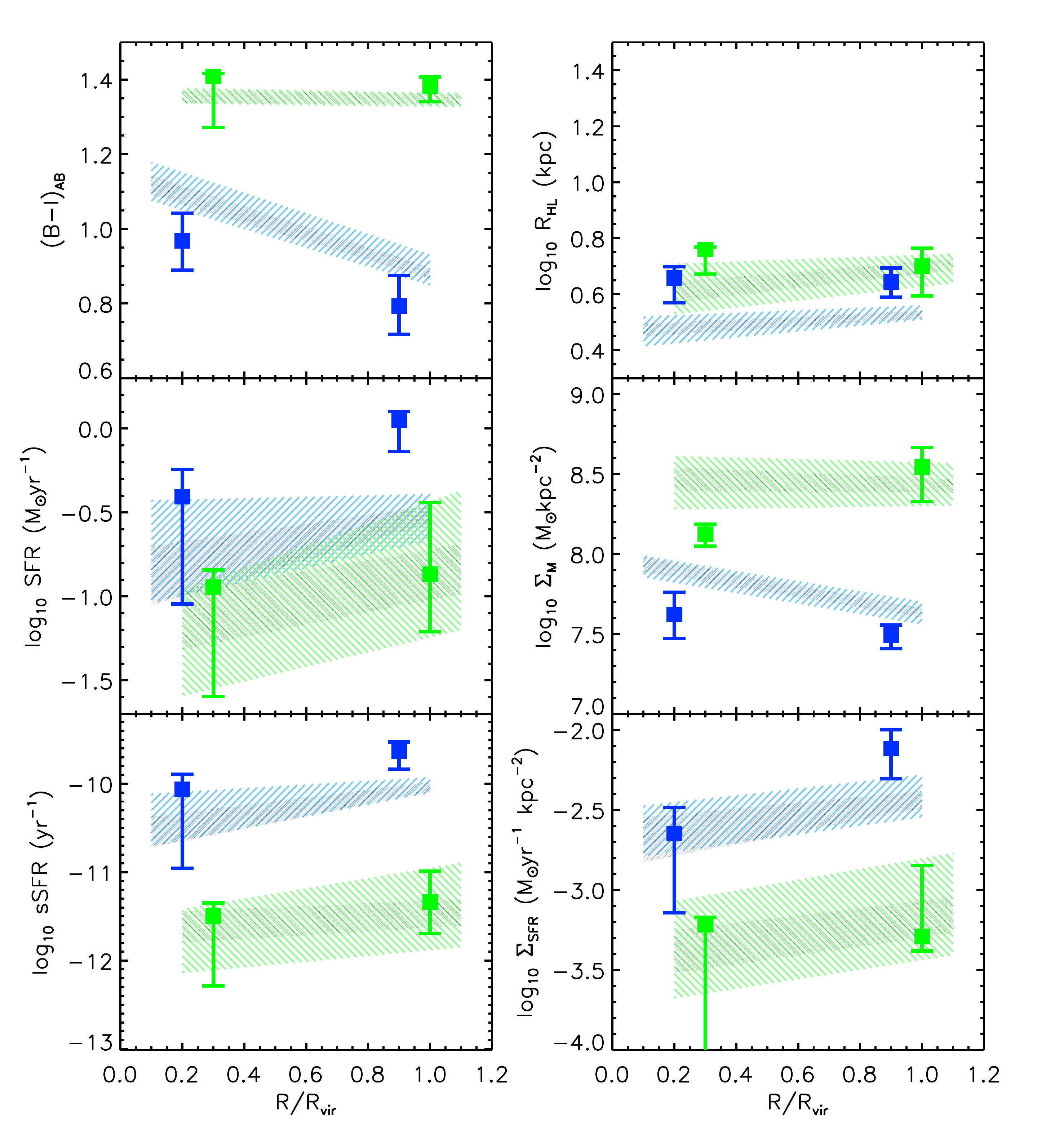}
\caption{Same as in Figure~\ref{Figure_8}, this time with merger properties at fixed $R/R_\mathrm{vir}$ (two bins at $R/R_\mathrm{vir}\le 0.5$ and $R/R_\mathrm{vir}> 0.5$).
 By definition, centrals are located at $R/R_\mathrm{vir}=0$ and are hence not plotted here.
\label{Figure_9}}
\end{figure*}

\begin{figure*}
\centering
\includegraphics[width=0.9\textwidth]{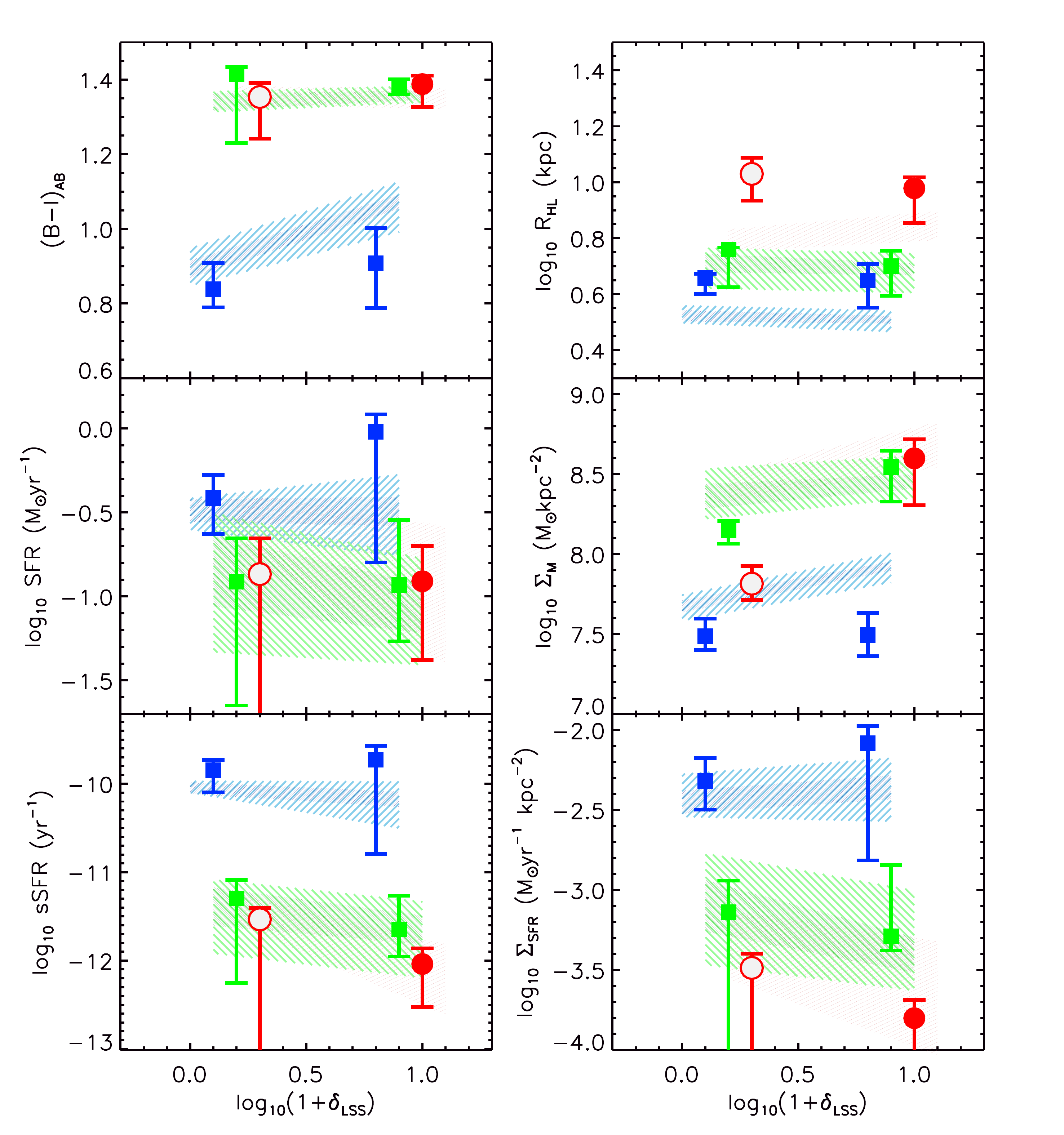}
\caption{Same as in Figure~\ref{Figure_8}, this time with merger properties at fixed LSS density (two environmental bins $\log(1+\delta_\mathrm{LSS})<0.7$ and $\log(1+\delta_\mathrm{LSS})>0.7$). 
\label{Figure_10}}
\end{figure*}

The environmental trends are highlighted in Figs.~\ref{Figure_8}-\ref{Figure_10} where we compare galaxies in mergers with control samples that are additionally drawn from the same (dense versus sparse) environments. In Figure~\ref{Figure_8} we focus on the role of $M_\mathrm{HALO}$, whereas in Figure~\ref{Figure_9} we split both merging and control samples into group-centric regions. Obviously, centrals have $R/R_\mathrm{vir}=0$ by construction hence they are not consider in this case. Finally, Figure~\ref{Figure_10} compares mergers and control samples at fixed LSS overdensities. In these figures, low-mass merging satellites are shown with blue dots and their control sample as a blue shaded area. Higher mass satellites are in green, and centrals are in red.

Within a {\it fixed} environmental bin, the results presented in the previous sections on the merging and control-sample satellites remain qualitatively unchanged, and we will therefore not repeat the discussion. We instead comment in the following on differential effects in the merger and control samples {\it across} the environmental bins; this may help us to understand the role of the environment in the transformations triggered by the merger event.

First, although we do not show this explicitly in a figure, we find that if we limit the analysis to relaxed groups the merger-induced enhancement in the SFR for systems with separation $<20$ kpc disappears because of a decrease in the median SFR of the mergers (as opposed to an increase in the median SFR of the control sample). In other words, a substantial fraction of the low-mass satellites in mergers with the highest SFRs are those in groups that are not yet relaxed. In contrast, low-mass satellites in mergers with low ($<1 M_{\odot} \mathrm{yr}^{-1}$) SFR are mostly in relaxed systems. We also mention the fact that the difference in radius at low galaxy masses becomes  insignificant when the study is limited to mergers (and galaxies in the control sample) in relaxed groups.

Second, the merger-induced variations in galaxy properties are present even if the given galaxy properties have themselves an environmental dependence. For instance, the median color of low-mass galaxies becomes bluer at lower halo masses and outer group radii (see Figure \ref{Figure_8} and \ref{Figure_9} and also Paper III). The relative change in the color with environment for the low mass merging satellites and the control sample is however of a similar amount, keeping the merger-induced bluening almost constant with environment. Therefore, turning the argument around, we demonstrate that the effect of mergers is to induce a $\sim 0.2$ mag blueing of the color irrespective of environment (within the errors) at low galaxy masses. This is a clear example that highlights the need for comparing merging galaxies and control samples at fixed mass \textit{and} environment in order to isolate the effect of mergers on galaxy properties. 

Third, for low-mass merging satellites, the enhancement in the SFR becomes more significant in the group outer regions. Although the radial trend is partly due to the fact that the low-mass nonmerging satellites are preferentially located at large radii rather than in the group cores thus reducing the scatter (i.e., the width of the shaded area) around the median SFR in the former regions, our data suggest that the {\it strength} of the merger-induced star-formation episode may change with environment. We checked, by generating random control subsamples matching in mass distribution and number of galaxies those involved in mergers in a given environmental bin, that any little difference in the mass distributions of both the merger and the control sample in the low galaxy mass bin did not contribute to the signal.

In conclusion, we are able to order the different environmental indicators by importance in setting the star-formation properties of the merging system: the most important indicator is the proximity to the companion, then the status (relaxed or unrelaxed) of the host group and finally the position within the halo.

\section{Physical Properties of Merging Centrals Relative to Equal-Mass Nonmerging Centrals} \label{sec:centrals}

We now focus on the properties of merging {\it central galaxies} with an analysis similar to that done for satellite galaxies. As highlighted in Section \ref{sec:biases}, this analysis of centrals is limited to the high galaxy mass bin, and is conducted at a median galaxy stellar mass of $10^{11} \Msol$. 

At these galaxy masses, merging central galaxies have red colors ($(B-I) \sim 1.4$ mag) similar to the control-sample nonmerging centrals of comparable masses, reflecting the fact that the high-mass centrals of our sample are mostly passively evolving systems, whether they are in a merger or not (middle and bottom panels of Figure~\ref{Figure_4}). There are exceptions, however. Among mergers involving central galaxies, our sample includes: $(1)$ one system with a high SFR $\sim 15~\Msol $yr$^{-1}$ in the secondary galaxy, i.e., at face value, a merger-induced burst in which a high SFR is triggered in the least massive galaxy; and $(2)$ two systems in which a highly enhanced SFR is observed, relative to the comparison nonmerging population, in the primary central galaxy (SFR$\sim 15~\Msol$ yr$^{-1}$).

Moreover, we note that high-mass secondaries (which are by definition satellite galaxies) that are members of merging pairs are as red as their primaries: the median color difference between primary central galaxies and high-mass secondary satellite galaxies involved in the merging pairs is always $(B-I)<0.1$ mag (Figure~\ref{Figure_4}). This is a straightforward consequence of the fact that both primary central galaxy and secondary satellite companion galaxies have stellar masses in the regime where passive bulge-dominated galaxies dominate the global galaxy population at $z=0$. Following a similar reasoning as in Section \ref{sec:satsizes}, the accreting primary (central) galaxy of the merging pair must have a very low gas-to-star fraction on average, motivating our classification of these mergers as dry in the discussion.

One-half of the centrals have companions in the lower galaxy mass bin. As a consequence of the stellar mass difference, these secondaries are bluer (by $\sim 0.3$ mag, median difference) than the primaries but their colors ($\sim 1.1$ mag, median value) and SFRs ($\sim 0.19~M_{\odot}$ yr$^{-1}$, median value) are consistent with those of their control sample. The median SFR in the centrals of such pairs are also low ($\sim 0.1~M_{\odot}$ yr$^{-1}$). Therefore these pairs would also result in relatively `dry' mergers.

A structural comparison of central galaxies involved in a merger with nonmerging central galaxies returns a marginally significant result in the direction of larger half-light radii in merging centrals by a factor of $\sim1.4$ in $d<20$ kpc pairs. This would formally yield surface mass densities in merging centrals about 0.3 dex lower on average than those in control sample centrals. With respect to the other properties considered here, merging centrals have negative color gradients and equally high S\'{e}rsic indices ($n\sim5$, Figure~\ref{Figure_6}, top panel) and bulge-to-total ratios (B/T=$0.45$, see Figure~\ref{Figure_6}) as in the control sample.

As is clear from Figs.~\ref{Figure_8} and \ref{Figure_10}, the environment (including the relaxed or unrelaxed status of the host group) does not strongly affect the stellar or structural properties of central galaxies involved in mergers. The only suggested environmental effect is an increase in the difference between the half-light radii of merging and noninteracting centrals with halo mass/LSS density, in the sense that merging centrals appear more noticeably extended than the control sample when located in a massive group or low $\delta_\mathrm{LSS}$. The number of ZENS central galaxies in these bins is however small ($N<5$), and the observed trends should be confirmed with larger samples.

\section{Discussion}\label{sec:discussion}

\subsection{How Do Mergers Affect the Properties of Galaxies in Groups?}

As highlighted in Sections \ref{sec:satsfr} and \ref{sec:centrals}, high galaxy mass mergers involving $>10^{10} M_\odot$ galaxies, whether centrals or satellites, are consistent with being mostly gas-depleted dry mergers. The result is not surprising in light of the environment-independent quenching of star formation in galaxies at these high masses \citep[e.g.,][]{peng10}, and it extends the findings of \citet{mcintosh08} to lower halo masses. Evidence for merger-induced structural changes in these quenched galaxies is minimal. A dry merger will lead to an increase in size of the coalesced remnant central galaxy, while roughly preserving its nuclear stellar velocity dispersion \citep[e.g.,][]{ciotti07}. This may lead to larger sizes in the central galaxy population than in the corresponding satellite population of similar galaxy stellar mass or velocity dispersion. This is convincingly observed in the galaxy size versus mass and galaxy size versus velocity dispersion relations for the brightest cluster members and satellite galaxies in galaxy clusters \citep[e.g.,][]{bernardi11}. 
Furthermore, although the observed size evolution of early-type galaxies can be largely explained by the appearance of newly quenched, extended galaxies with cosmic time, some degree of size growth for individual galaxies is expected at the highest masses (see \citealt{carollo13a} for a discussion on both effects).
Mergers of massive central galaxies with quenched a satellite, like those observed in our high-mass sample, could be a potential channel for size variations in these individual objects.

At low galaxy masses, there is plenty of merging activity among star-forming satellites.\footnote{We remind readers that we do not attempt a study of passive mergers at these low galaxy masses because our sample is not complete for such systems below $10^{10} M_\odot$ and that only satellites populate the ZENS groups at these galaxy mass scales.}
 Moreover, at variance with the high-mass case, the low-mass mergers can induce star formation and merging satellites below $10^{10.2} M_\odot$ exhibit a factor of $\sim 2-3$ enhancement in SFRs and sSFRs relative to their nonmerging counterparts. Noticeably, the fact that such an effect is evident only at low separations ($d<$20kpc) reinforces the notion that mergers in their final stages profoundly alter the behavior of galaxies, in the very least in terms of boosting their rate of consumption of their gas reservoirs \citep[see also, e.g.,][]{larson78, kennicutt87, lambas03, solalonso06, woods07, dimatteo08, kampczyk13}. In particular, simulations show that any enhancement in the SFR generally occurs either at the first pericenter or at the coalescence, depending on the orbit configuration \citep[e.g.,][]{dimatteo07} and if both the eccentricity and the impact parameter are high, there might be an intermediate phase during which the star formation is suppressed \citep{patton13}.

Large catalogs of simulated mergers show that the strongest bursts happen in a minority (15\%) of the cases and are short lived, whereas 76\% of fly-bys and 50\% of mergers display an integrated SFR only 1.25 larger than that of isolated galaxies (\citealt{dimatteo07, dimatteo08}, see also \citealt{woods10, patton13}). This is consistent with our findings. The predicted star-formation enhancement in these simulations is comparable during fly-bys and coalescence, justifying a posteriori our choice to discuss mergers at coalescence and those in very close pairs at the same time. Numerical simulations further show that such mergers often stop gas accretion onto the coalesced remnant \citep{feldmann11}. As mentioned in Section \ref{sec:satsfr}, however, the $\la1.5$ Gyr mass-doubling timescales of merging satellites in our sample imply that these systems will possibly reach an undisturbed appearance in the coalesced remnant well before their current gas reservoir is exhausted. 
This is potentially interesting in light of the results in  \citet{carollo14}, where we discuss the possibility that quenching may not significantly alter the structural properties of galaxies, which are instead set prior to star-formation cessation, and find that the increase in the median B/T of quenched satellite galaxies with respect to star-forming ones can be explained by fading of the disks following quenching.

Despite the boost in sSFR induced by the merger event, the rate of surface star formation per kpc$^2$ remains similar at the $\Sigma_{SFR} \sim 10^{-2.3}~\Msol $yr$^{-1} $kpc$^{-2}$ level, both for merging and nonmerging low-mass satellites, on account of their different sizes (i.e., 50$\%$ larger in the interacting satellites). These may indicate that feedback is strong enough that it prevents star formation from concentrating in the central galactic regions, leading to an increase in measured galaxy sizes relative to the progenitors \citep[e.g.,][]{hopkins08}. Moreover, a more diffuse distribution of star formation seems to emerge in simulations with a sufficient physical resolution to better reproduce the real distribution of star formation within galaxies \citep[e.g.,][]{teyssier10}. Similar evidence for a reduced gas flow toward the galaxy cores during interactions (relative to earlier pioneering experiments, e.g., \citealt{mihos94}) is also found in other recent simulations \citep{cox08, dimatteo07}, which indicate that feedback is most likely responsible for preventing gas from flowing to the center and feeding a centrally concentrated burst \citep[see also][]{hopkins13, newton13}. Our data offer global support for this scenario.

\subsection{Which of the Different Environments Leads to Galaxy Mergers?}

From an environmental perspective, our data suggest two main factors that either preferentially induce or preferably enable galaxy mergers: either galaxies inhabit the relatively small potential wells of low-mass group halos (see Figure~\ref{Figure_1}) or  they lie at the bottom of the potential wells of their host groups, i.e.,  they are central galaxies (see Figure~\ref{Figure_2} and the discussion in Section \ref{sec:GroupDistance}).

The fact that, at any galactic mass in our study, galaxy mergers are favored at low halo masses is consistent with the expectation based on simple dynamical friction timescale arguments of satellites falling into groups \citep{chandrasekhar43, binney87}, as well as with analytical estimates that take into account the merger cross section for direct mergers \citep[e.g.,][]{mamon92, mamon00, makino97}. That is potentials with relatively low velocity dispersions are more conducive to galaxy--galaxy interactions than more massive, higher velocities halos \citep[see also][]{perez09b, mcintosh08, heiderman09, tal09}. Halo-occupation models further support this interpretation by showing that the merger efficiency (i.e., the merger timescale relative to the Hubble time) is about an order of magnitude higher at halo masses $M_\mathrm{HALO}\sim10^{12.5}~M_{\odot}$ than in massive halos with $M_\mathrm{HALO}>10^{13.5}~M_{\odot}$ \citep{hopkins08}. 

These overall trends discussed when showing Figure~\ref{Figure_1} aggregate both mergers onto the central and satellite--satellite mergers. At high galaxy masses, from Figure~\ref{Figure_1} we can infer that the biggest contribution to the slope $\Delta \Gamma \over \Delta \log M_\mathrm{HALO}$ comes from mergers involving central galaxies. This is further highlighted in Figure~\ref{Figure_11}, where we present once again our results on the merger fraction as a function of $M_\mathrm{HALO}$ for relaxed groups, this time considering separately mergers with a central as the primary galaxy (left panel) and mergers with a satellite as the primary galaxy (right panel).
The slope $\Delta \Gamma \over \Delta \log M_\mathrm{HALO}$ stays almost constant with pair separation around a value of $\simeq -0.07$. Such a value for the slope can be qualitatively understood in terms of the merger rate scaling as the inverse of the group velocity dispersion to the third power \citep[e.g.,][]{mamon00}, further considering that this latter quantity scales as $M_\mathrm{HALO}^{0.3}$ \citep[e.g.,][]{Bryan_Norman_1998}, and finally taking into account that the merger timescale (necessary to convert the merger rate into fractions) seems to be independent of the halo mass \citep{kitzbichler08}. 

At high galaxy masses, the predominance of central--satellite mergers in our sample is also in very good agreement with previous studies based on independent samples. For example, in their SDSS sample of $\ga 10^{11} M_\odot$ galaxies, \citet{mcintosh08} find that at least half of their merger events involve a central--satellite systems. The increase in the merger rate with decreasing group-centric distance is expected in analytical estimates \citep[e.g.,][]{mamon00}. Moreover, the important role of mergers in building up the central galaxies of groups is well documented in numerical simulations \citep[e.g.,][]{feldmann10, feldmann11}, and is supported by observations at both similar  \citep{tal09, rasmussen10} and earlier epochs \citep[e.g.,][]{rines07, tran08, lin10}  as that of the ZENS sample.

Finally, we note that the mild increase in the merger fraction of low galaxy mass satellites in regions of low (relative to high) LSS density is also found in other data sets \citep[see, e.g.,][]{ellison10}. This would also be supported by the predictions of the \citet{guo11} semianalytical model that show a decrease (9\% $\rightarrow$ 6\%) in the merger fraction as the density increases, in agreement within the errors with the ZENS data.

\subsection{Comparison with Semianalytical Models}

In order to make a more quantitative comparison with theoretical expectations, we made use of the publicly available\footnote{Available at http://gavo.mpa-garching.mpg.de/Millennium/ \citep{lemson06}.} data from the \citet{guo11} semianalytical model. In brief, the model follows the dynamics of the subhaloes hosting the satellite galaxies until their dark matter content exceeds their baryonic mass, switching a \textit{merger clock} based on the estimated dynamical friction timescale afterward. By construction, satellite galaxies merge with the central of their halo. Groups fulfilling the ZENS specifics in terms of redshift, masses, number of members and $b_J$ magnitude of the galaxies were selected from a snapshot of the simulation box that was projected onto one of its axes in order to define the projected LSS density exactly as in Paper I and to compute projected distances among galaxies. Mergers were then identified according to the projected separation and velocities $<500\rm km s^{-1}$, as in the empirical sample. 

\begin{figure*}
\includegraphics[width=\textwidth]{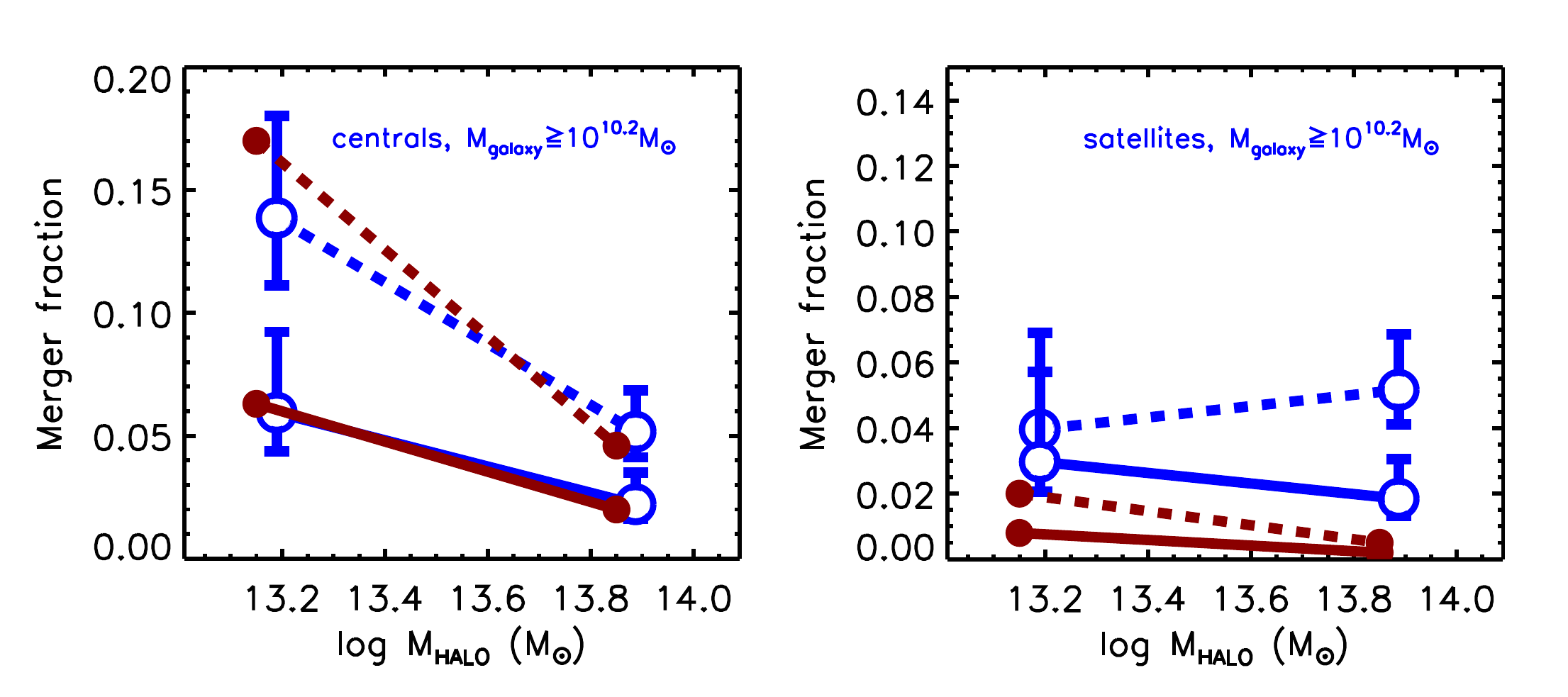}
\caption{Merger fraction for pairs as a function of halo mass for the high galaxy mass bin and {\it relaxed groups only}. Close ($d<20$ kpc) pairs are shown with solid lines and dashed lines are for all pairs ($d<50$ kpc). The mass of the primary (most massive) galaxy is used to place each pair in the suitable galaxy mass bin. Left: the primary galaxy is also the central galaxy of the host group.  Right: the primary galaxy is a satellite galaxy. Predictions from the semianalytical model by \citet{guo11} for high-mass mergers are displayed as dark red points and connected by lines with the same coding as a function of separation. We find that mergers involving high-mass satellites as primaries are significantly underestimated in the model. 
\label{Figure_11}}
\end{figure*}

The dark red lines in Figure~\ref{Figure_11} show the predictions from \citet{guo11} for the merger fraction versus $M_\mathrm{HALO}$ relation. As shown in the left panel, by selecting $d<$20 kpc ($<$50 kpc) pairs involving central galaxies, we estimate in the models a merger fraction of $\sim$6\% ($\sim$17\%) at low $M_\mathrm{HALO}$ and $\sim$2\% ($\sim$5\%) at higher group mass, respectively. The predicted variation $\Delta\ \Gamma/\Delta \log (M_\mathrm{HALO})$ of the merger fraction with halo mass for mergers involving a central is in excellent agreement with our findings at small separations. When selecting $<$50 kpc pairs, the model predicts instead a steeper slope than the observed one, but still in agreement within the errors.

Because in the \citet{guo11} model the satellite mergers are computed on the basis of the dynamical friction formulae, in light of what was discussed at the outset of this section, it is not surprising that the simulation broadly matches the observed trend with halo mass. It is, however, remarkable the agreement in the normalization, namely in the actual merger fractions. Given this consistency with observations for the merger fraction in groups of different halo masses, we can use the model predictions to estimate that 1/6 ($\sim$1/2) of the central--satellite mergers at low (high) halo masses and d$<$20 kpc in ZENS will coalesce in less than $\sim $300Myr, whereas the remainder of the systems will take $\sim$1Gyr.

The right panel of Figure~\ref{Figure_11} shows that mergers involving only high-mass satellite as primaries are instead significantly underestimated in the model with respect to our measure for relaxed groups. 
In the \citet{guo11} semianalytical model, real satellite--satellite mergers are in general extremely rare, with a fraction lower than 0.1\% that is well below our measured fraction. In the model, these mergers involve a central of a halo that has been accreted and still is a cooling site. Similarly rare are satellite--satellite mergers fraction in other semianalytical models \citep[e.g.,][]{pipino09}. 

There are two possible interpretations of these findings. The first is that these models miss a key  ingredient, namely satellite--satellite merging,  which, if happening in nature (as indicated by our data), would be an important channel through which environment operates on galaxies. Another possibility is, however, that our results are affected by projection effects, which increase the number of close satellite--satellite pairs. To disentangle between these alternatives, we  estimate the fraction of close ($d<20(50)$ kpc) low-mass satellite--satellite pairs that are sufficiently distant from their central galaxy to avoid satellite--central mergers; we thus set a threshold distance of 50 kpc between the satellite--satellite pair and the central galaxy.  In the models, the fraction of such systems that we find in low and high $M_\mathrm{HALO}$ groups is respectively  $\sim0.2\ (1.2)$\%  and $\sim0.1\ (0.5)$\%. These fractions are well below  our estimated fractions in the ZENS sample  (see Figure~\ref{Figure_1}), and they provide support for the interpretation that only a small fraction of our detected satellite--satellite mergers are due to projection effects that lead to galaxy closeness or even superposition.

Dark matter simulations also provide support in the same direction. \citet{angulo09} show that satellite--satellite mergers occur in a $\Lambda$CDM universe and become increasingly more frequent with decreasing subhalo versus\, main-halo mass ratio. These authors interpret this result as being due to mergers of progenitor satellites that are members of a subhalo that is in the process of being accreted by a larger halo. 

We finally also use the models to estimate whether spurious galaxy superpositions may have led us to overestimated the central--satellite merger fractions. To this purpose we compute in the ZENS sample the fraction of non-member galaxies which lie at a projected distance $<20(50)$ kpc from the central of a given group. We compare this fraction with the population of  galaxies in the simulation cube that lie within a 20 Mpc thick slice centered on the given group, which have a velocity difference $<500~\rm km~s^{-1}$ and projected distance $<3 R_\mathrm{vir}$ relative to the central galaxy of that group. This returns a probability of a random projection $<0.1\%$ at all group masses under study. We therefore also conclude that the central--satellite merger fractions that we have estimated from our sample are not severely affected by superposition effects.

\subsection{An Empirical View of the Effects of Mergers on Galaxy Evolution}

In contrast to the above-mentioned complex galaxy formation simulations, a simple yet successful empirical picture of galaxy formation has been recently presented to explain the role of galaxy mass and environment in the evolution of the mass function of both active and passive galaxies \citep{peng10}.

This scenario explains the evolution of galaxies as a population driven by the cosmic run of the sSFR and its interplay with both internally and environmentally induced quenching rates. For what concerns our paper, it is useful to remind that in the \citet{peng10} framework, mergers are proportionally more important for the growth of massive galaxies than at lower galaxy masses, and that the merger fraction is about a factor of four higher in the densest regions (which in their definition roughly corresponds to the inner group regions, \citealt{peng12}) than in the less dense environment. Massive galaxies, however, have already been \textit{mass} quenched. Therefore, the mergers involving massive galaxies are dry and the fraction of passive galaxies undergoing subsequent dry mergers quickly increases at masses above $10^{11}~M_{\odot}$. 

This is consistent with ZENS data where: (1) the median mass of centrals involved in high mass mergers exceeds $10^{11}~M_{\odot}$; (2) their colors are as red as galaxies in the control sample at the same mass, and (3) a strong radial dependence of the merger fraction linked to the growth of the centrals (e.g., Figure~\ref{Figure_2}) is found. In addition, we find that the star formation is generally not enhanced in low-mass satellites merging with centrals. 

Taking into account the mass ratios of the pairs in our sample, we expect that the increase in mass caused by these dry mergers will be between 15\% (median contribution of low-mass secondaries) and 38\% (median contribution if the secondary is in the high galaxy mass bin). A quantitatively similar average increase in mass is also expected by \citet{peng10} considering the constraints given by the galaxy mass function evolution. Therefore, our data offer to the broad picture of galaxy formation a quantitative estimate of the typical growth of central galaxies in the $z\sim 0.05$ group environment.

\section{Summary and concluding remarks}\label{sec:summ}

We have utilized the sample of $0.05<z<0.0585$ ZENS galaxies with $M>10^{9.2}\Msol$ (1274 galaxies) and its 162 identified merging or close pairs systems to investigate: $(1)$ the dependence of the merger fraction $\Gamma$ on three different measurements of environment, i.e., halo mass, group-centric distance, and LSS density; and $(2)$ the internal properties of merging satellite and central galaxies in comparison with galaxies of a control sample having similar stellar masses, rank within the host group potentials, and environment. Our main findings are as follows:

\begin{enumerate}
\item In relaxed groups and at any galaxy mass scale in the range $10^{9.2} M_\odot$-- 10$^{11.7} M_\odot$, an  enhancement in $\Gamma$ by a factor of $\sim 3$ is observed in groups with masses $M_\mathrm{HALO}<10^{13.5}~M_{\odot}$ relative to higher mass groups. The sharp drop in $\Gamma$ at halo masses $M_\mathrm{HALO}> 10^{13.5}~M_{\odot}$ suggests that merger activity is effectively suppressed by the large galaxy velocities sustained in high mass halos. In the case of relaxed groups, we infer a variation in the merger fraction $\Delta\ \Gamma/\Delta \log (M_\mathrm{HALO}) \sim - 0.07$ dex$^{-1}$, which is almost independent on galaxy mass and merger stage. 

\item A similar increase in the merger fraction at low LSS densities ($\log_{10}(1+\delta_\mathrm{LSS})<0.7$) is seen  in our data at the $\gtrsim2.5 \sigma$ level for low-mass pairs or high-mass mergers involving a central galaxy. We find that it holds at the $\sim2 \sigma$ level if we restrict the analysis to groups with $M_\mathrm{HALO}<10^{13.7}~M_{\odot}$ and hence it is not caused by an underlying correlation between halo mass and overdensity.  This interesting suggested effect is thus a possible indication of the influence of the LSS density on low-mass mergers and central--satellites interactions which should be further investigated with larger data sets.

\item Most mergers that we observe in our sample at galaxy masses $>10^{10.2} M_\odot$ are central--satellite, gas-depleted dry mergers \citep[see also, e.g.,][]{edwards12}. These high-mass dry mergers occur preferentially in relaxed groups, and support numerical experiments indicating a substantial role of mergers in the mass assembly histories of the central galaxies of group halos \citep[e.g.,][]{delucia07}. 

\item The high frequency of central--satellite mergers at high galaxy masses results in a nominal increase of $\Gamma$ in the inner, $R<0.5R_\mathrm{vir}$ group regions; this highlights the need to accurately disentangle the galaxy populations of groups/clusters into centrals and satellites, in order to properly interpret any observational trend in galaxy properties with group or cluster-centric distance.

\item At low galaxy masses of $<10^{10.2} M_\odot$, where in ZENS we are only probing satellite galaxies, we find evidence of merger-induced star formation in the gas-rich (satellite--satellite) mergers. 
 The relevant environments in determining the amount of induced star formation are, in decreasing order of importance, the proximity to the companion, the status (relaxed/unrelaxed) of the host group, and to a minor extent the position within the halo. In particular, the low-mass satellite--satellite mergers have (specific) SFRs enhanced by a factor of $\sim 2-3$ relative to their nonmerging counterparts when we consider either coalesced or very close ($<20$ kpc) systems. Furthermore, a substantial fraction of the low-mass satellites in mergers with the highest SFR are those in groups that are not yet relaxed, whereas low-mass satellites in mergers with low ($<1 M_{\odot} yr^{-1}$) SFR are mostly in relaxed systems. 

\item From the analysis of the color maps and the comparison of the color gradients of merging galaxies to those of the control sample, it appears that these boosted star-formation activities are mostly distributed over the whole galaxy body, rather than being concentrated in a nuclear `burst' of star formation. The most recent numerical simulations also correct earlier claims for a merger-driven nuclear enhancement of star formation and ascribe to stellar feedback the diffusion of the star-formation activity throughout the extent of the galaxy \citep{cox08}. This may also explain the $\sim1.5\times$ larger sizes (implying $\sim$2--3$\times$ lower surface mass densities) of these low-mass merging satellites relative to nonmerging satellites of similar mass. 

\item The mass doubling timescales for low-mass satellite--satellite mergers is 2--3 Gyr, a factor of $\sim 3$ times shorter than that of similar nonmerging galaxies, but it remains nevertheless a factor of $\sim3$ longer than numerically estimated merging timescales for these systems \citep{cox06}. 

\end{enumerate}

\section*{acknowledgments}
We acknowledge the support from the Swiss National Science Foundation. This publication makes use of data from ESO Large Program 177.A-0680. We also use data products from the Two Micron All Sky Survey, which is a joint project of the University of Massachusetts and the Infrared Processing and Analysis Center/California Institute of Technology, funded by the National Aeronautics and Space Administration and the National Science Foundation. This research has made use of the NASA/IPAC Extragalactic Database (NED) which is operated by the Jet Propulsion Laboratory, California Institute of Technology, under contract with the National Aeronautics and Space Administration. The Millennium Simulation databases used in this paper and the web application providing online access to them were constructed as part of the activities of the German Astrophysical Virtual Observatory (GAVO). A.P. thanks P. Di Matteo for useful comments on numerical simulations.

\end{document}